\documentclass[Afour,sageh,times]{sagej}
\usepackage{moreverb,url}
\usepackage[colorlinks,bookmarksopen,bookmarksnumbered,citecolor=red,urlcolor=red]{hyperref}
\newcommand\BibTeX{{\rmfamily B\kern-.05em \textsc{i\kern-.025em b}\kern-.08em
T\kern-.1667em\lower.7ex\hbox{E}\kern-.125emX}}
\def\volumeyear{2021}

\usepackage{subcaption}
\usepackage{algorithm, algorithmic}

\begin{document}

\runninghead{Einkemmer and Moriggl}

\title{Semi-Lagrangian 4d, 5d, and 6d kinetic plasma simulation on large scale GPU equipped supercomputers}
\author{Lukas Einkemmer\affilnum{1} and Alexander Moriggl\affilnum{1}}

\affiliation{\affilnum{1}University Of Innsbruck, Austria}

\corrauth{Alexander Moriggl,
University of Innsbruck
Techniker.~13,
A-6020 Innsbruck,
Austria}

\email{alexander.moriggl@uibk.ac.at}

\begin{abstract}
Running kinetic plasma physics simulations using grid-based solvers is very demanding both in terms of memory as well as computational cost. This is primarily due to the up to six-dimensional phase space and the associated unfavorable scaling of the computational cost as a function of grid spacing (often termed the curse of dimensionality). In this paper, we present 4d, 5d, and 6d simulations of the Vlasov--Poisson equation with a split-step semi-Lagrangian discontinuous Galerkin scheme on graphic processing units (GPUs). The local communication pattern of this method allows an efficient implementation on large-scale GPU-based systems and emphasizes the importance of considering algorithmic and high-performance computing aspects in unison. We demonstrate a single node performance above 2 TB/s effective memory bandwidth (on a node with 4 A100 GPUs) and show excellent scaling (parallel efficiency between 30\% and 67\%) for up to 1536 A100 GPUs on JUWELS Booster.
\end{abstract}

\maketitle

\section{Introduction}

Computational fluid dynamics has become an extremely successful framework to understand a range of natural phenomena ranging from astrophysics to everyday technological applications. Despite these successes, it is increasingly realized that many phenomena in plasma physics require a kinetic description \cite{howes2008,zweibel2009}. The most challenging aspect of kinetic models, from a computational point of view, is certainly the up to six-dimensional phase space. The traditional approach to overcome this difficulty has been to employ so-called particle methods \cite{verboncoeur2005,Hariri2016pic}. While these simulations are comparably cheap and have been successfully employed in some applications, it is well known that they miss or do not resolve accurately certain physical phenomena, such as Landau damping. In addition, the inherent numerical noise introduced makes it extremely difficult to obtain accurate results for many physical phenomena. More recently, sparse grids \cite{kormann2016,sparseGrid2016} and low-rank methods \cite{einkemmer2018low,Einkemmer2021lr} have been used to reduce memory requirements and computational complexity. Nevertheless, the need to run fully kinetic simulations in high dimension persists for a range of applications.

To run such simulations large supercomputers are needed and codes to facilitate this for CPU based systems are available. See, e.g., \cite{bigot2013scaling} for a scaling study of the 5-dimensional GYSELA code. More recently, also simulations for the full 6 dimensional problem have been considered \cite{Kormann2019}. In the latter work scaling to up to 65,536 processes has been reported on the SuperMUC HPC system of the Leibnitz
Supercomputing Center. We also note that the amount of communication required usually increases as we increase the dimensionality of the problem. The reason for this is that the ratio of boundary points (that need to be transferred) to interior grid points increases significantly.

To numerically discretize these models, semi-Lagrangian schemes have proved popular. By using information on the characteristics of the underlying partial differential equations (PDEs), these methods do not suffer from the usual step size restriction of explicit methods. A particular emphasize in collisionless models, as we will consider here, is to introduce as little numerical diffusion as possible into the approximation. It is well known that semi-Lagrangian schemes based on spline interpolation (see, e.g., \cite{sonnendrucker1999semi} and \cite{Filbet2003interpolations} for a comparison of different methods) and spectral methods (see, e.g., \cite{klimas1994,klimas2018,camporeale2016velocity}) perform well according to this metric. 

However, the results for scaling to large supercomputers obtained in \cite{Kormann2019} use Lagrange interpolation instead. The reason for this is that in a massively parallel environment, the global communication patterns induced by spline interpolation or spectral methods provide a significant impediment to scalability. This, in particular, is also a serious issue on GPU-based systems (even on a single node; see, e.g., \cite{Einkemmer2020GPUs}). Recently, it has been demonstrated that a different type of semi-Lagrangian scheme, the local semi-Lagrangian discontinuous Galerkin approach, can be implemented efficiently on both Tesla and consumer GPUs, while maintaining or even exceeding the performance of spline-based semi-Lagrangian schemes \cite{Einkemmer2020GPUs}.

In this paper, we present results, using our SLDG code, that demonstrates that large-scale kinetic simulations of the Vlasov--Poisson equation can be made to scale efficiently on modern GPU-based supercomputers. This is particularly relevant as most (if not all) pre-exascale and exascale systems will make use of GPUs and kinetic simulations are prime candidates that could exploit such systems. 


\section{Problem and algorithmic background}
\label{sec:nummethods}

\subsection{Vlasov--Poisson equation}

To describe the kinetic dynamics of a plasma we use the Vlasov--Poisson equations
\begin{align}
 &  \partial_t f + v\cdot \nabla_x f - E(f)\cdot \nabla_v f = 0, \label{eq:vlasov} \\
 & E = -\nabla_x \phi , \quad -\Delta \phi = 1 - \rho, \quad \rho = \int f\,dv, \label{eq:poisson}
\end{align}
where $f(t,x,v)$, with $(x,v) \in \mathbb{R}^{d_x} \times \mathbb{R}^{d_v}$, is the (up to) six-dimensional particle density in phase space, $\rho(t,x)$ the (up to) three-dimensional charge density and $E(t,x)$ is the (up to) three-dimensional electric field. The Vlasov--Poisson equation is a hyperbolic model that couples a transport equation self-consistently to an equation for the electric field. The electric field is determined via a Poisson problem from the charge density $\rho$.

The particle density is up to six-dimensional. Depending on the physical model numerical simulations can be run using different number of dimensions in space (denoted by $d_x$) and in velocity (denoted by $d_v$). In this paper we consider problems with 2x2v ($d_x=2$ and $d_v=2$, four dimensional), 2x3v ($d_x=2$ and $d_v=3$, five dimensional), and 3x3v ($d_x=3$ and $d_v=3$, six dimensional).

While we restrict ourselves to the Vlasov--Poisson model in this work, we emphasize that the performance results obtained provide a good indication for other kinetic models as well. Both on the plasma physics side (such as in case of the Vlasov--Maxwell equations) as well as for kinetic problems from radiative transport.

\subsection{Time splitting}

The first step in our algorithm for solving the Vlasov--Poisson equation is to perform a time splitting. The approach we use here has been introduced in the seminal paper \cite{CHENG1976} and has subsequently been generalized to, e.g., the Vlasov--Poisson equation \cite{crouseilles2015hamiltonian}.

The idea of the splitting approach is to decompose the nonlinear Vlasov--Poisson equation into a sequence of simpler and linear steps. The strategy to compute the numerical solution at time $t^{n+1} = t^n + \Delta t$, where we use $f^n \approx f(t_n,x,v)$, is as follows
\begin{itemize}
\item[1.] Solve $ \partial_t f(t,x,v) + v\cdot \nabla_x f(t,x,v) = 0 $ with initial value $f(0,x,v) = f^n(x,v)$ to obtain $f^{\star}(x,v)=f(\Delta t/2, x, v)$.
\item[2.] Compute $E$ using $f^{\star}$ by solving the Poisson problem \eqref{eq:poisson}.
\item[3.] Solve $\partial_t f(t,x,v) - E(x) \cdot \nabla_v f(t,x,v)=0 $ with initial value $f(0,x,v) = f^{\star}(x,v)$ to obtain $f^{\star\star}(x,v) = f(\Delta t,x,v)$.
\item[4.] Solve $\partial_t f(t,x,v) + v\cdot \nabla_x f(t,x,v) = 0$ with initial value $f(0,x,v) = f^{\star\star}(t,x,v)$ to obtain $f^{n+1}(x,v) = f(\Delta t/2,x,v)$.
\end{itemize}
This results in a second-order scheme (i.e. Strang splitting). 
Since the advection speed never depends on the direction of the transport, the characteristics can be determined analytically. For example, for step 1. we have
\[ f^{\star}(x,v) = f^n(x-v \Delta t/2,v). \]
This advection in $d_x$ dimensions can be further reduced to one-dimensional advections by the following splitting procedure
\begin{itemize}
  \item[1a.] $f^{\star a}(x_1,x_2,x_3,v) = f^n(x_1-v_1 \Delta t/2,x_2,x_3,v)$.
  \item[1b.] $f^{\star b}(x_1,x_2,x_3,v) = f^{\star a}(x_1,x_2-v_2 \Delta t/2,x_3,v)$.
  \item[1c.] $f^{\star}(x_1,x_2,x_3,v) = f^{\star b}(x_1,x_2,x_3-v_3 \Delta t/2,v)$.
\end{itemize}
We emphasize that no error is introduced by this splitting as the corresponding operators commute. We apply a similar procedure to step 3, which results in the following
\begin{itemize}
 \item[3a.] $f^{\star a}(x,v_1,v_2,v_3) = f^n(x,v_1 +E_1(x)\Delta t,v_2,v_3)$.
 \item[3b.] $f^{\star b}(x,v_1,v_2,v_3) = f^{\star a}(x,v_1,v_2+E_2(x)\Delta t,v_3)$.
 \item[3c.] $f^{\star}(x,v_1,v_2,v_3) = f^{\star b}(x,v_1,v_2,v_3+E_3(x)\Delta t)$.
\end{itemize}
Since the advection in the $v$-direction (step 3a-3c) does not change the electric field, we have reduced the up to six-dimensional Vlasov--Poisson equation to a sequence of one-dimensional linear advections and a single Poisson solve. The scheme outlined here is second-order accurate \cite{einkemmer2014a}. High-order methods have been constructed as well \cite{casas2017high} and generalization of this basic approach to more complicated problems also exist (see, e.g., \cite{crouseilles2015hamiltonian} for the Vlasov--Maxwell equations).

The advections are treated using a so-called semi-Lagrangian method, which evaluates the function approximation at the foot of the characteristic curves. Since these end points of the characteristics do not lie on the grid points, an interpolation technique has to be applied. We will describe this procedure in detail in the next section.

\subsection{Semi-Lagrangian discontinuous Galerkin method}

Semi-Lagrangian methods have been widely used to solve the Vlasov equation. In particular, spline interpolation \cite{sonnendrucker1999semi,filbet2001conservative} has emerged as a popular method as it does not introduce too much numerical diffusion, is conservative, and gives good accuracy at a reasonable computational cost.

However, in the construction of a spline each degree of freedom couples with each other degree of freedom. Due to this global data dependency, spline interpolation is not suited to distributed memory parallelism. To remedy this in \cite{crouseilles2009parallel} a method that performs spline interpolation on patches has been proposed. However, on today's massively parallel systems with relatively few degrees of freedom per direction on each MPI process (especially in the six-dimensional case), it is unclear how much advantage this approach provides. In this context, it should be noted that the massively parallel implementation found in \cite{Kormann2019} forgoes using spline interpolation and focuses exclusively on Lagrange interpolation. The same is true for the GPU based implementation in \cite{mehrenberger2013vlasov}. While Lagrange interpolation scales well it is also known to be very diffusive and usually large stencils are required in order to obtain accurate results.

In the present work, we will use a semi-Lagrangian discontinuous Galerkin scheme (sldg). This method, see e.g.~\cite{crouseilles2011discontinuous,rossmanith2011}, is conservative by construction \cite{einkemmer2017study}, has similar or better properties with respect to limiting diffusion compared to spline interpolation (see \cite{Einkemmer20194d} for a comparison), and is completely local (at most two neighboring cells are accessed to update a cell). Thus, it provides a method that in addition to being ideally suited for today's massively parallel CPU and GPU systems, is competitive from a numerical point of view. In addition, a well-developed convergence analysis is available for the Vlasov--Poisson equation
\cite{einkemmer2014a,einkemmer2014b}.

\begin{figure*}[h]
\centering
 \includegraphics[width=1.0\textwidth]{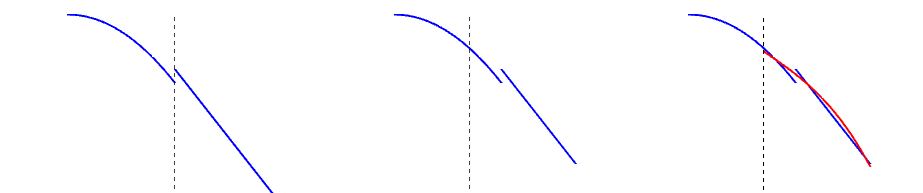}
 \caption{\label{fig:dg}Illustrates the semi-Lagrangian discontinuous Galerkin method. When the advection is complete (central picture), a projection has to be performed which results in the red line in the right picture. This projection lies in the approximation space and is used for subsequent computations. Note that only two adjacent cells are required to compute the values for a certain cell.}
\end{figure*}

The sldg scheme computes in an efficient way $u^{n+1}(y) = u^n(y-at)$. The one-dimensional advections in the previous section are written precisely in this form (all the remaining variables are considered as parameters here). This scheme first divides the domain into cells. The restriction of the solution to each such cell is a polynomial of degree $k$. Note that there is no continuity constraint across cell interfaces and the approximant is thus discontinuous across this interface (see the left picture in figure~\ref{fig:dg}). The basis to form these polynomials are Lagrange polynomials which interpolate at the Gauss--Legendre points scaled to each cell. Thus, the degrees of freedom are function values $u_{ij}^n$ where the index $i$ is the cell index and the index $j$ refers to the $j^{th}$ Gaus--Legendre point of cell $i$. Since the translated function $u^{n+1}$ in general does not lie in this approximation space, an $L^2$ projection is applied, see figure~\ref{fig:dg}. Only two adjacent cells are required in order to compute the values at a given cell, we denote the index of them with $i^*$ and $i^*+1$. The resulting numerical scheme has order $k+1$ and can be written as follows
\begin{equation}
 u^{n+1}_{ij}=\sum_lA_{jl}u^n_{i^*l} + \sum_l B_{jl}u^n_{i^*+1;l}.
 \label{eq:dgmat}
\end{equation}
The matrices $A$ and $B$ are of size $(k+1) \times (k+1)$ and only depend on the advection speed, which can be precomputed. More details about the specific form of these two matrices can be found in \cite{Einkemmer20194d}.

\subsection{Poisson solver}
To determine the electric field, the Poisson equation~\eqref{eq:poisson} has to be solved once in each time step. However, since this is only an up to three-dimensional problem, the computational cost of this part of the algorithm is small. Using FFT based methods is a common approach that is also supported in our SLDG code. 

When the spatial variables $x$ are partitioned over multiple compute nodes, however, the FFT method requires global communication. Thus, we propose an alternative method which requires only local communication to solve this problem. Since the advection is already treated using a discontinuous Galerkin scheme, we will use the same approximation space for the Poisson solver. This has the added benefit that no interpolation to and from the equidistant grid on which FFT based methods operate is necessary. The details of the discontinuous Galerkin Poisson solver are outlined in the Appendix.

\section{Implementation and parallelization \label{sec:implementation}}

Our code is written in C++/CUDA and uses templates to treat problems with different number of dimensions in physical and velocity space. This allows us to only maintain one codebase,  while still providing excellent performance in all these configurations. Our code SLDG is available at \url{https://bitbucket.org/leinkemmer/sldg} under the terms of the MIT license.

The main computational effort of the code is due to the following three parts:

\textbf{Advection:} In the advection step equation~\eqref{eq:dgmat} is applied to each cell.This is done in a CUDA kernel and the operation is largely memory bound on modern GPUs. This step is required whether we run the code on a single GPU or on a large supercomputer and its cost per degree of freedom is expected to be independent of the number of MPI processes taking part in the simulation.

\textbf{Communication:} In order to run the code on multiple GPUs that possible span multiple nodes a domain decomposition approach is used. Depending on the problem, either only the velocity direction or both the space and velocity directions are divided along $d_x$ or $d_x+d_v$ directions. Each of these blocks is assigned to a single GPU. Performing an advection step necessitates the exchange of boundary data from GPU to GPU. On a single node this can be done using a call to \texttt{cudaMemcpy}, which makes use of NVLink if available. Communication between nodes is done via MPI. Our implementation supports the use of CUDA-Aware MPI which can make use of GPU memory in MPI calls. Thus, in this configuration MPI routines can directly read from and place data in GPU memory.

\begin{figure}[h]
  \centering
  \includegraphics[width=0.4\textwidth]{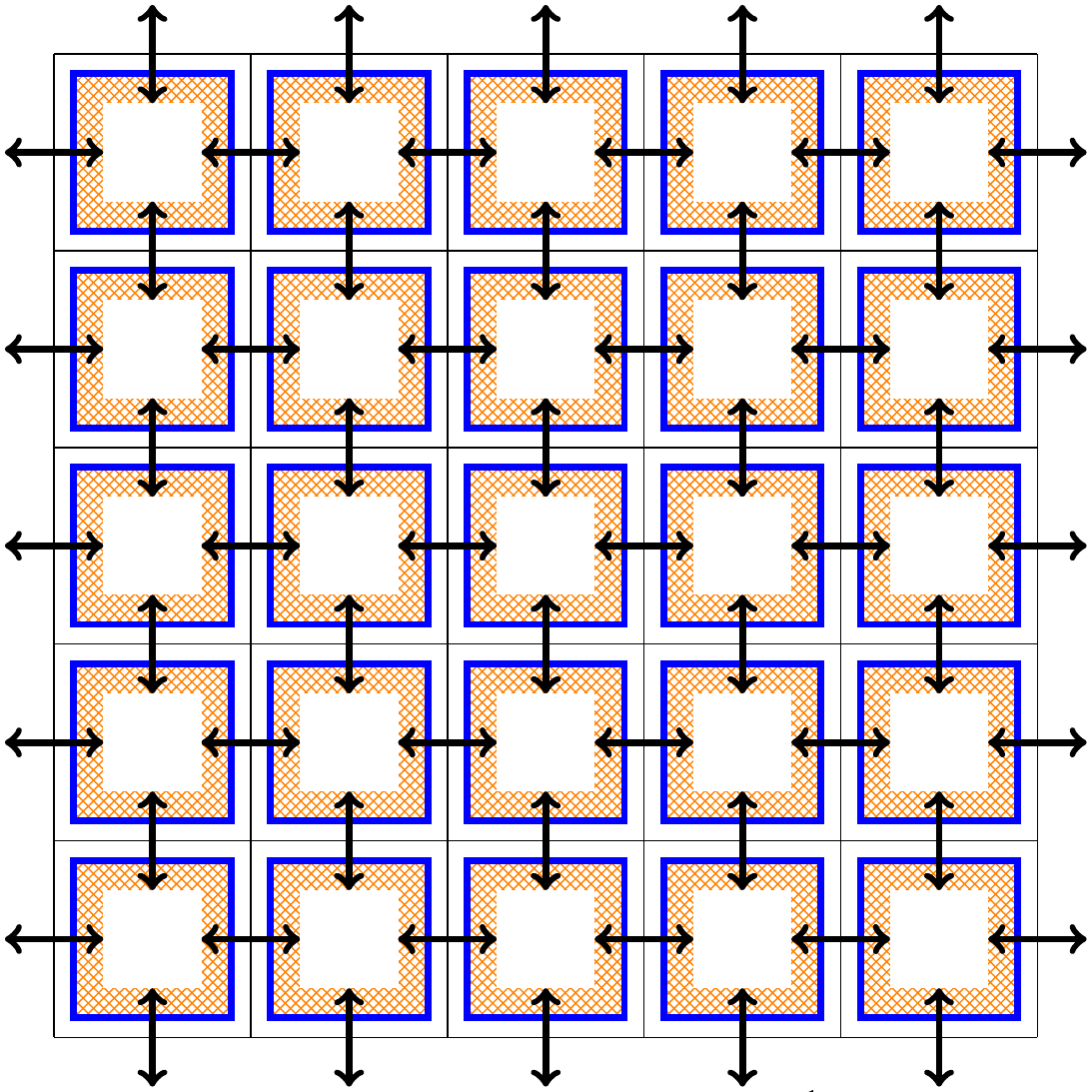}
    \caption{Illustration of the data transfer between GPUs in the 1+1 dimensional case (i.e.~in the $x-v$ plane.The color orange represents the ghost cell data that potentially (depending on the direction of the velocity) has to be sent to neighboring GPUs. Black arrows indicate the corresponding data transfer }
    \label{fig:global_transfer}
\end{figure}

\begin{figure}[h]
     \centering
     \includegraphics[width=0.4\textwidth]{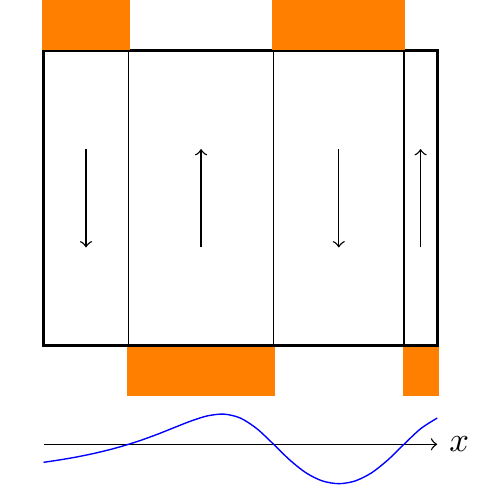}
     \caption{Demonstration of the advection in the velocity direction and the required boundary data (orange blocks) from neighboring processes. The direction of data transfer depends on the sign of the velocity (the blue function below the illustration).}
     \label{fig:data_transfer_demo}
\end{figure}

Especially for high-dimensional problems, this type of boundary exchange can be relatively expensive. The reason for this is that the number of degrees of freedom in each spatial and velocity direction that can be stored on a single GPU is relatively small (compared to a three-dimensional problem). Thus, boundary cells constitute a larger fraction of the overall degrees of freedom, which due to the discrepancy between GPU memory and InfiniBand bandwidth means that communication can take up a significant fraction of the total run time. The data transfer is illustrated in figure~\ref{fig:global_transfer}. Let us emphasize, however, that not the entire boundary of each data block associated to a GPU has to be transferred to each neighboring GPU. Since advection is directional, depending on the sign of the velocity field only either a send or a receive is required. This is illustrated in figure~\ref{fig:data_transfer_demo}. By taking this fact into account, only half the boundary data has to be transferred.

\begin{figure}[h]
  \centering
  \includegraphics[width=0.4\textwidth]{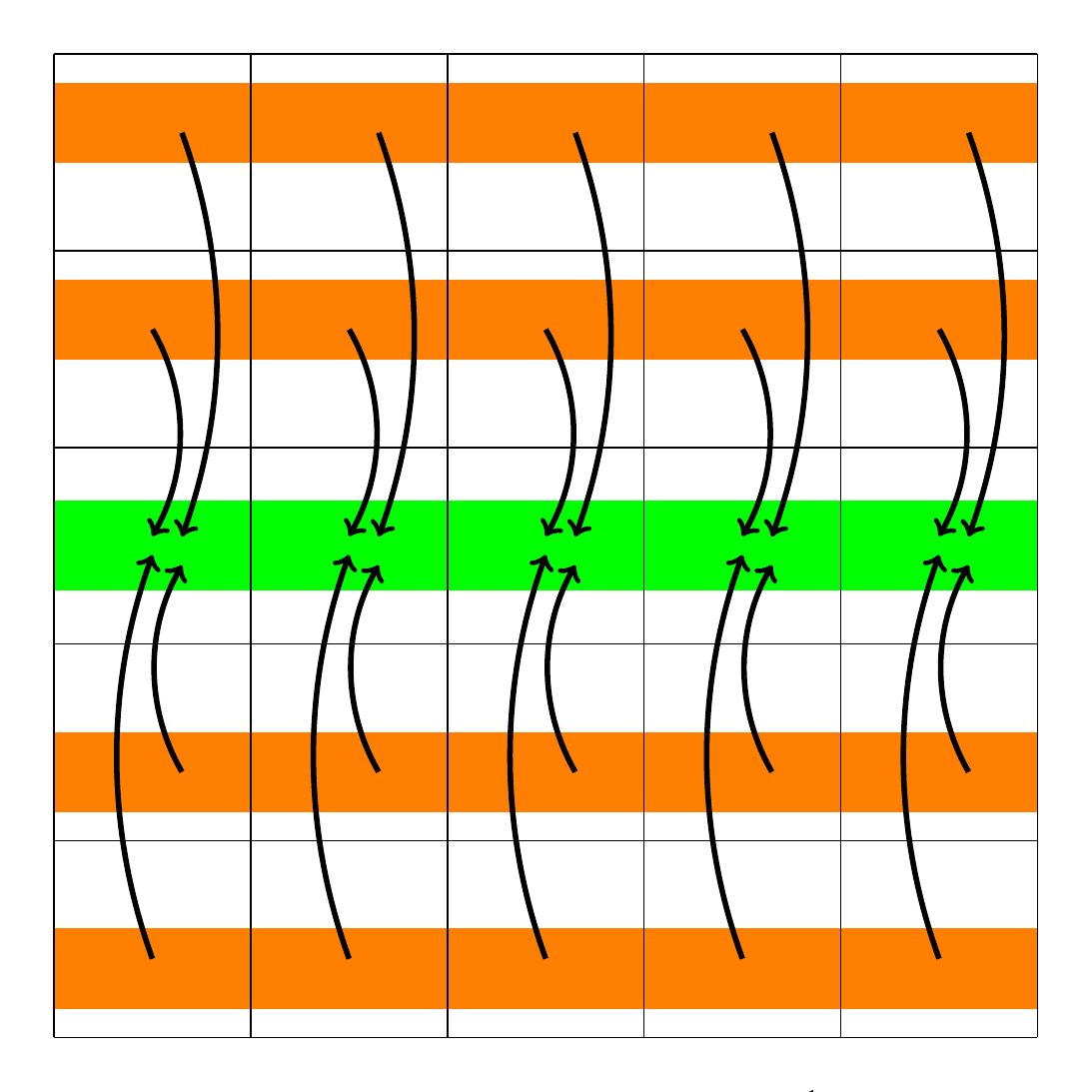}
  \caption{Illustration of the data transfer necessary to compute the charge density $\rho$ (Step 1 in computing the E field).}
  \label{fig:reduction}
\end{figure}

\textbf{E field:} The computation of the electric field proceeds in the following steps. First, the charge density $\rho(x) = \int f(x,v) \,dv$ has to be determined. This is first done on each GPU (using the CUB library \cite{CUB}) and then a reduction operation in the velocity direction is performed by calling \texttt{MPI\_Reduce} on processes which span the entire velocity domain but represent the same part of the spatial domain (as illustrated in Figure \ref{fig:reduction}). 

\begin{figure}[h]
  \centering
  \includegraphics[width=0.4\textwidth]{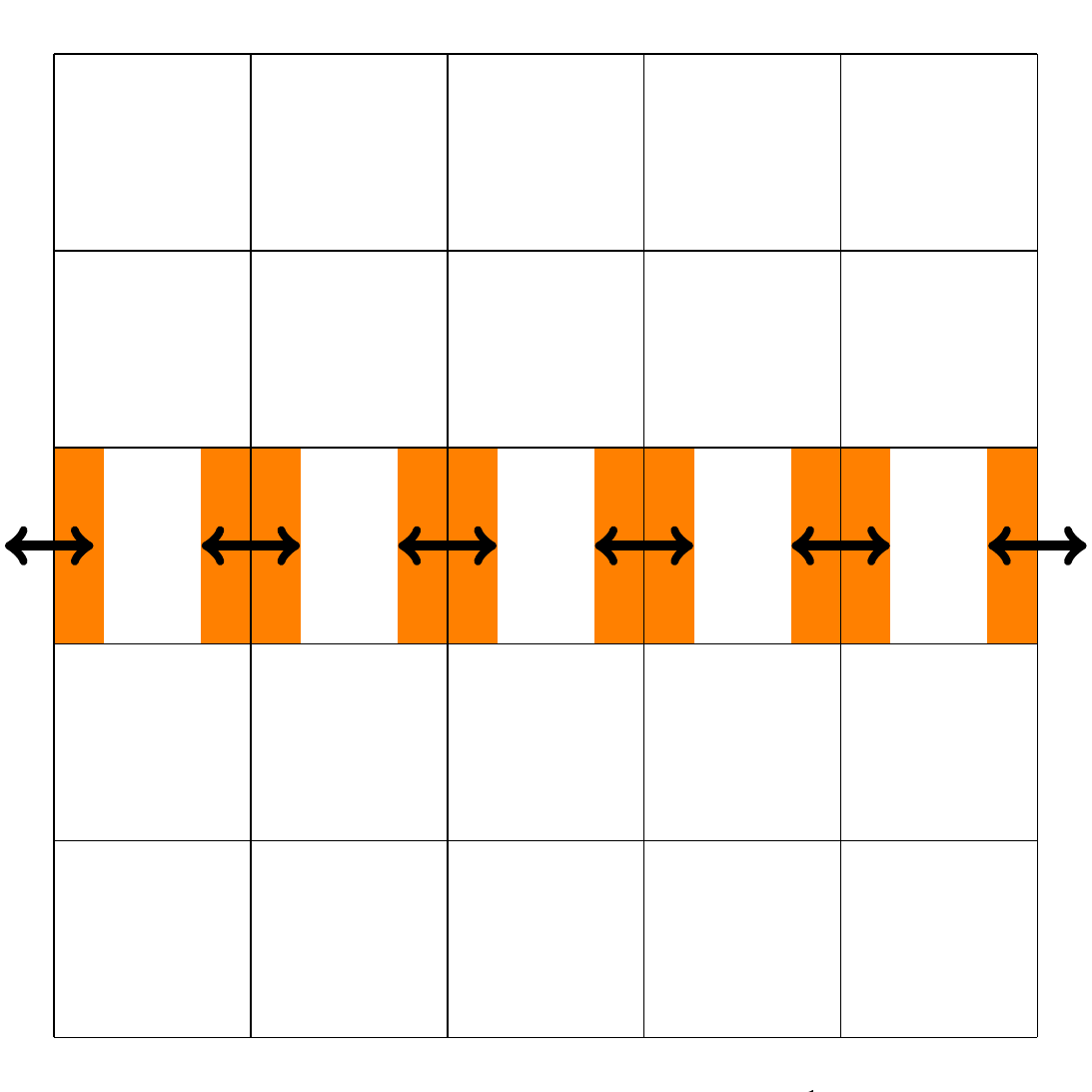}
  \caption{Illustration of the data transfer required to compute the electric field from the charge density $\rho$ (Step 2 in computing the E field).}
  \label{fig:poisson_mpi}
\end{figure}

At the completion of this step $\rho$ is distributed over certain MPI root processes which span the entire space domain. These processes then participate in computing the electric field $E$. This can be done either by an FFT based method, which requires global communication among the processes, or an iterative conjugate gradient discontinuous Galerkin (CGdG) solver which only requires nearest neighbor exchange in each iteration (as illustrated in Figure \ref{fig:poisson_mpi}).

\begin{figure}[h]
  \centering
  \includegraphics[width=0.4\textwidth]{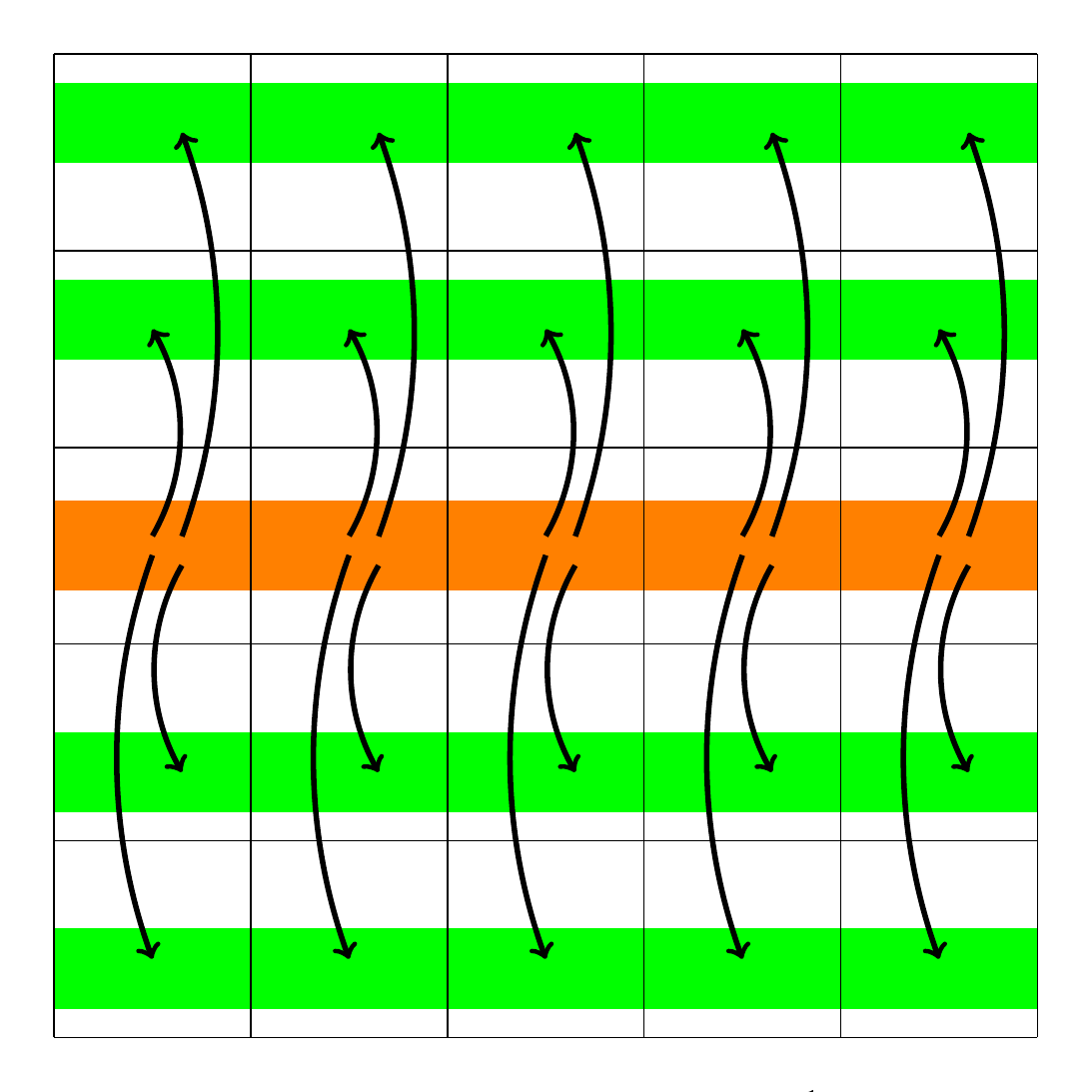}
  \caption{Illustration of the data transfer required to distribute the electric field to the appropriate processes (Step 3 in computing the E field).}
  \label{fig:broadcast}
\end{figure}

Finally, the electric field has to be made available to all processes that span the velocity domain. To do this \texttt{MPI\_Broadcast} is called, see figure \ref{fig:broadcast}.

Although from a computational perspective the computation of the electric field is relatively cheap (as it is only an up to three-dimensional problem), the associated data transfer can become an impediment to scaling, especially on large systems.

In a multi-node context each MPI process is assigned one or multiple blocks. Each block manages a single GPU. The configuration we employ on most clusters with multiple GPUs on each node is to have one MPI process per node (i.e.~one MPI process per shared memory domain) and consequently as many blocks per MPI process as GPUs are on a node. However, on Juwels Booster this significantly degrades the bandwidth that is available to conduct MPI communication between different nodes. Thus, the multi-node simulations in this paper are run using a single MPI process per GPU. On Juwels Booster therefore 4 MPI processes are launched on each node and CUDA-aware MPI is used both for inter-node as well as intra-node communication.

\subsection{Validation}
\label{sec:validation}

To validate our implementation, two commonly used test scenarios are considered. The linear Landau damping problem given by the initial value
\[
 f(0,x,v) = \frac{\exp\left(-\Vert v \Vert^2/2\right)}{(2\pi)^{d_v/2}}\left(1+\alpha \sum_{i=1}^{d_x} \cos(k x_i)\right)
\]
on $[0,4\pi]^{d_x}\times [-6,6]^{d_v}$, $\alpha=10^{-2}$, and $k=0.5$ and a bump-on-tail instability
\begin{align*}
  & f(0,x,v) = \\
 &\frac{1}{(2\pi)^{d_v/2}}\left(0.9\exp\left(-\frac{v_1^2}{2}\right) + 0.2\exp\left(-2(v_1-4.5)^2\right)\right) \\
 &\exp\left(-\frac{\sum_{i=2}^{d_v}v_i^2}{2}\right)\left(1+0.03\sum_{i=1}^{d_x}\cos(0.3x_i)\right).
\end{align*}
on $[0,20\pi /3]^{d_x}\times[-9,9]^{d_v}$.

\begin{figure}[h]
  \centering
  \includegraphics[width=1.0\linewidth]{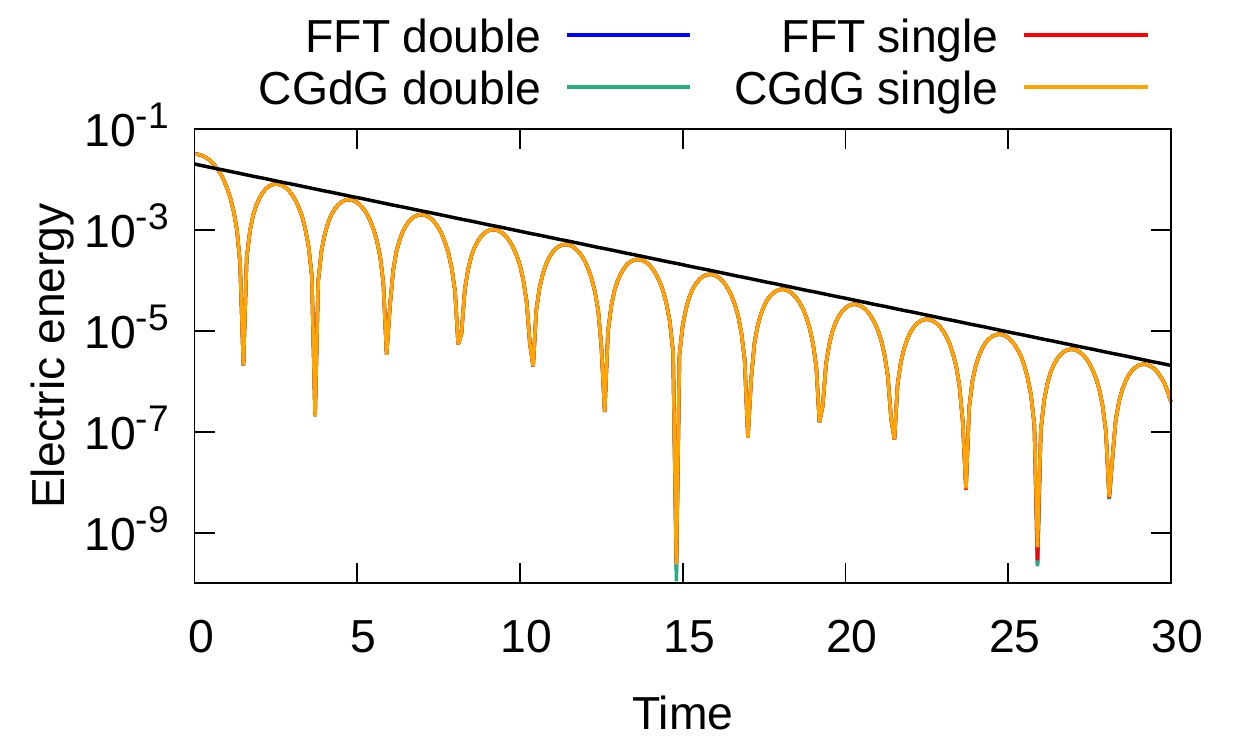}

  \includegraphics[width=1.0\linewidth]{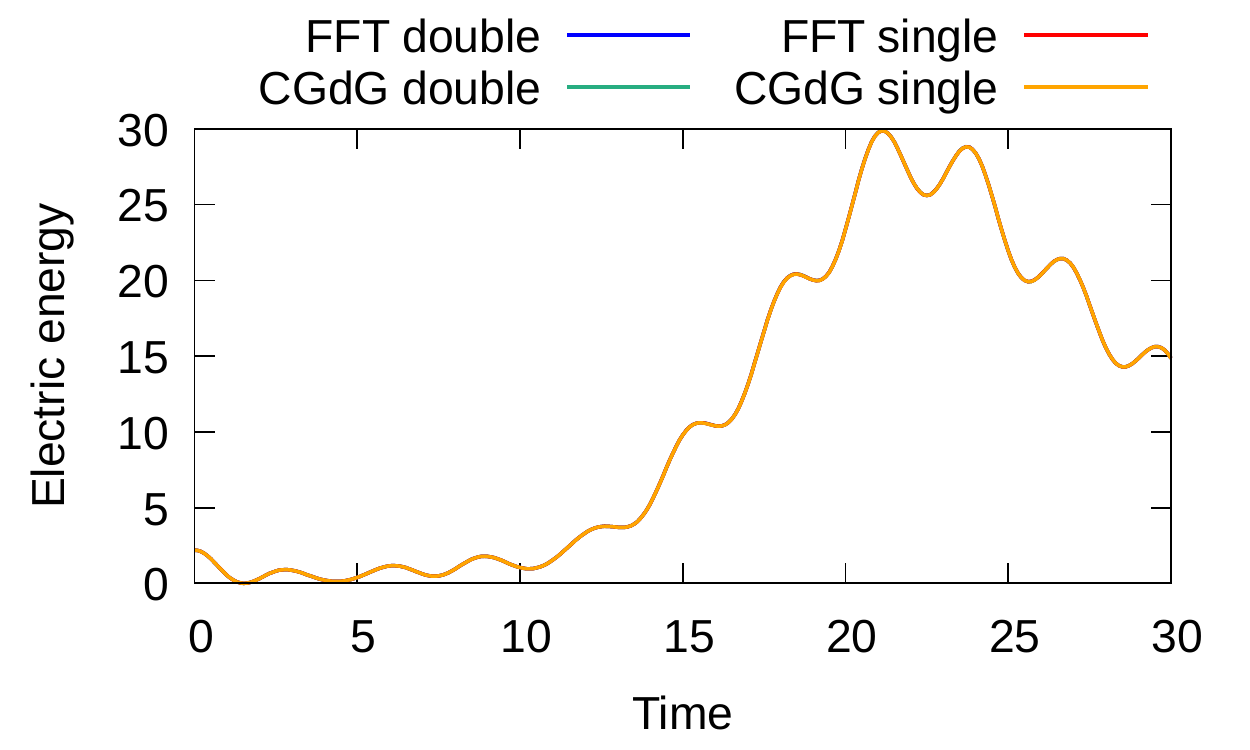}
\caption{\label{fig:5dllbot} Time evolution of the electric energy for the linear Landau damping (top) and bump-on-tail instability (bottom) for the 2x3v case. For both test cases the $4^{\text{th}}$ order method with a time step of $0.1$ is used. For the linear Landau test case $72^2144^3$ degrees of freedom are used (the domain is distributed over 8 GPUs), while for the bump-on-tail test scenario $144^5$ degrees of freedom are used (the domain is distributed over 32 GPUs).}
\end{figure}

\begin{figure}[h]
  \centering
  \includegraphics[width=1.0\linewidth]{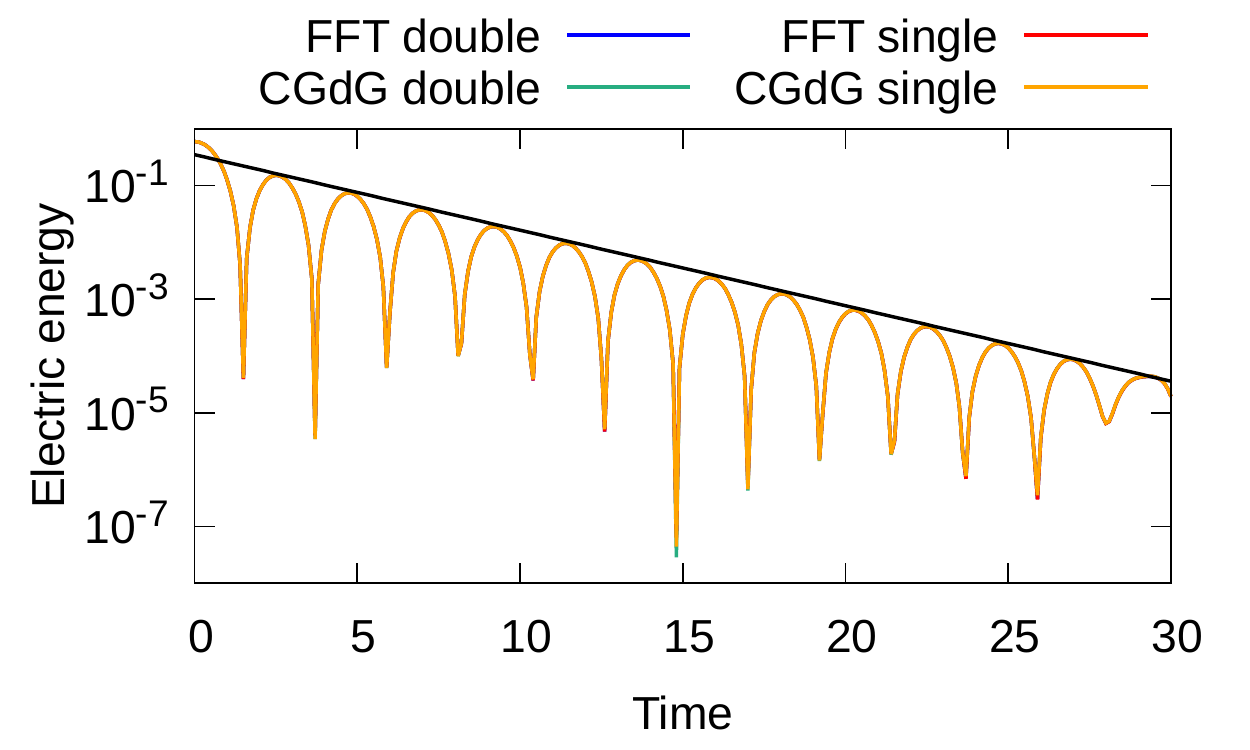}

  \includegraphics[width=1.0\linewidth]{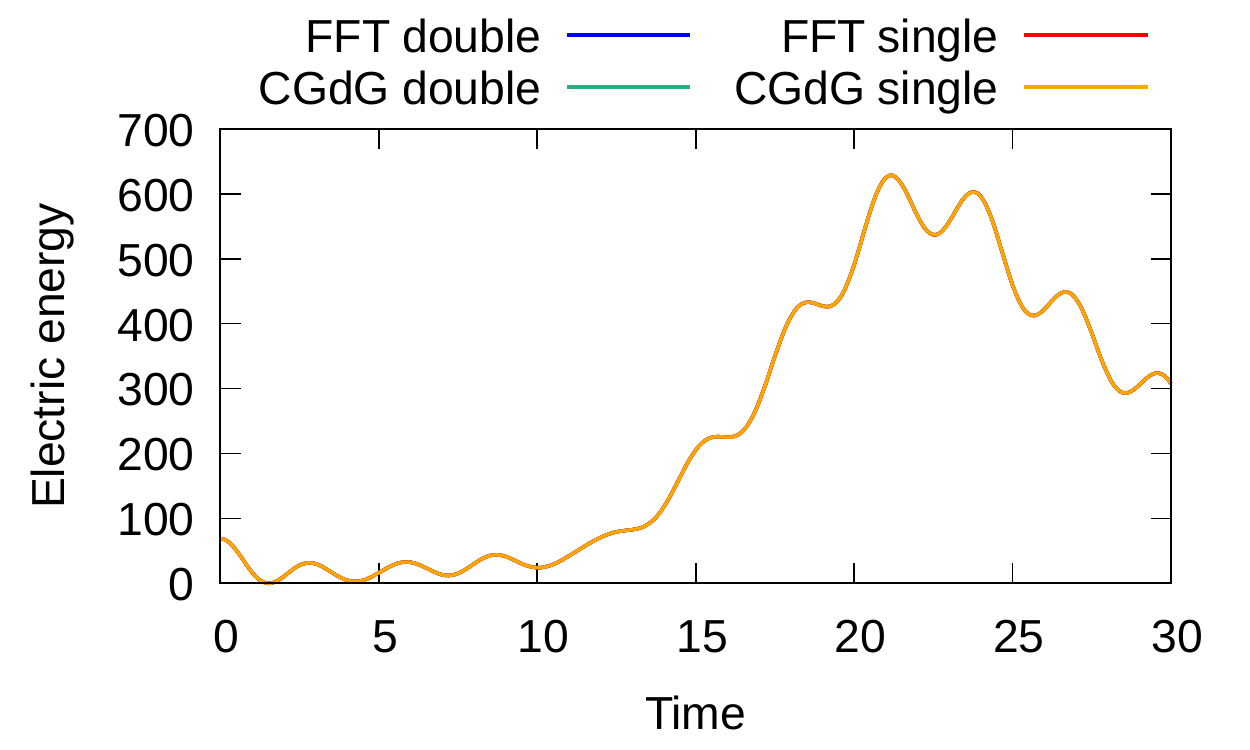}
\caption{\label{fig:6dllbot} Time evolution of the electric energy of the linear Landau damping (top) and bump-on-tail instability (bottom) for the 3x3v case. For both test case the $4^{\text{th}}$ order method with a time step size of $0.1$ is used. For the linear Landau test case $36^372^3$ degrees of freedom are used (the domain is distributed over 8 GPUs), while for the bump-on-tail test scenario $72^6$ degrees of freedom are used (the domain is distributed over 64 GPUs).} 
\end{figure}

In the simulations presented here we use the $4^{\text{th}}$ order semi-Lagrangian discontinuous Galerkin scheme (as it represents a good compromise between accuracy and computational cost \cite{einkemmer2017study,Einkemmer2020GPUs}) and a time step size of $0.1$. In figures~\ref{fig:5dllbot} and~\ref{fig:6dllbot} the time evolution of the electric energy for these initial conditions is shown for a five and six-dimensional configuration, respectively. The results agree well with what has been reported in the literature. In addition, the Landau damping results show excellent agreement with the analytically obtained decay rate. There is no significant difference between the results obtained with the FFT and discontinuous Galerkin Poisson solver. The SLDG code also includes a set of unit tests that check a range of results against known analytic and numerical solutions and compares the results obtain on the GPU to the results on the CPU.

We further remark that the results obtained using single and double precision overlap in the plots. Thus if one is not interested in too stringent tolerances running simulations with the semi-Lagrangian discontinuous Galerkin scheme in single precision is a viable option. This reduces memory consumption by a factor of two. In particular, for GPUs this can be attractive as memory, in general, is more limited on GPU based systems. Thus, larger problem sizes can be treated per GPU. In addition, for some problems consumer cards can potentially be used, which do not have strong double-precision performance. We will present and discuss both double and single precision results in the remainder of this paper.

\section{Single node performance}
\label{sec:performance}

The results in this section are obtained simulating the linear Landau test case. If not otherwise stated, a fixed time step size of $0.1$ and the $4^\text{th}$ order semi-Lagrangian discontinuous Galerkin method is used. To measure the performance of our code, the achieved memory bandwidth (BW) is used as a metric. This allows us, for our primarily memory bound problem, to easily compare the performance of our implementation with the peak performance of the hardware we run the simulations on. In addition, the run time can be easily obtained from this metric as well.

For the 3x3v (i.e.~the six-dimensional) problem solved using the second-order Strang splitting scheme nine advections have to be performed in each time step. This implies that the entire data has to be accessed 18 times (one read and one write per advection). For the 2x2v and 2x3v problem, 6 advections have to be performed and the data is consequently accessed 12 times. Moreover, for calculating the electric field once more the entire data has to be accessed (to compute the density $\rho$). This is done twice (once at the end of the time step to compute diagnostic values such as the electric energy and once at the middle of the time step). Therefore, we have
\[
 \text{achieved bandwidth} = \frac{F_{d_x,d_v} \cdot \text{sizeof}(\texttt{fp}) \cdot \text{dof} \cdot 10^{-9}}{\text{time per step in seconds}} \text{GB/s}
\]
where \texttt{fp} is \texttt{single} or \texttt{double} and $F_{d_x,d_v}$ is $20$ for the 3x3v problem and 14 otherwise. For comparison, we give in Table~\ref{tab:gpus} the theoretical peak bandwidth for the different GPUs that we will use in this work. These are the values specified by the vendors and the actual bandwidth obtained, even in a simple benchmark such as copy, is somewhat less.

\begin{table}[h]
\centering
 \begin{tabular}{l|r|r}
  GPU & memory bandwidth & memory size \\ \hline
  GTX1080 & 320 GB/s &  8 GB \\
  Titan V & 653 GB/s & 12 GB \\
  V100    & 900 GB/s & 16 GB \\
  A100    & 1\,555 GB/s & 40 GB \\
 \end{tabular}
\caption{The different GPUs used in this work and their specifications.}
\label{tab:gpus}
\end{table}

\subsection{Single GPU performance}

First, we compare the performance of the FFT and discontinuous Galerkin Poisson solver. The results are presented in table~\ref{tab:fftvsdg}. The run time for the FFT and the discontinuous Galerkin method is very similar, particular for the 2x3v and 3x3v problem. For the 2x2v problem the FFT based scheme is faster by approximately 30\%. This can be explained by the fact that the degrees of freedom per direction are reduced for higher dimensional problems (due to computational constraints) and thus not as many iterations are needed to obtain convergence. Since the FFT Poisson solver is faster and there is no disadvantage of using it for single node computations (we will discuss its role in scaling to larger systems later in this work), all of the computations performed in this section use the FFT based Poisson solver. 

\begin{table}[h]
\centering¸
 \begin{tabular}{c|c|c|c|c}
& & \multicolumn{1}{c}{FFT} & \multicolumn{2}{|c}{CGdG} \\ 
& dof & time/step  & time/step & Average iter.\\ \hline 
2x2v & $220^4$ & 0.372s & 0.477s & 173 \\
2x3v & $72^5$ & 0.299s & 0.306s & 55   \\
3x3v & $36^6$ & 0.619s & 0.646s & 38
 \end{tabular}
\caption{\label{tab:fftvsdg}Performance using different Poisson solvers on our local GPU cluster. The computations are done in double precision on an A100 GPU.}
\end{table}

In table \ref{tab:gpubandwidth5} and \ref{tab:gpubandwidth6} we show the achieved bandwidth in the 2x3v and the 3x3v case using four different types of GPUs. We achieve approximately half the bandwidth of what is theoretically possible on each platform. We consider this an excellent result as the memory access pattern is significantly less uniform than in a stencil code and there is a significant amount of computation that is not taken into account by our estimate of the achieved bandwidth (e.g.~computing the electric field from the density $\rho$; see \cite{Einkemmer2020GPUs} for a more detailed discussion).
\begin{table}
 \centering
\begin{tabular}{c|c|c|c|c|c}
    & GTX & TitanV & V100 & \multicolumn{2}{c}{A100} \\ 
dof & single & single & single & single & double \\ \hline
 $56^5$ & 104 & 356 & 412 & 583 & 649 \\
 $60^5$ & 100 & 377 & 430 & 607 & 670\\
 $64^5$ &     & 401 & 482 & 666 & 717\\
 $68^5$ &     &     & 454 & 634 & 696\\
 $72^5$ &     &     &     & 680 & 727\\
 $76^5$ &     &     &     & 660 & \\
 $80^5$ &     &     &     & 708 & \\
 $84^5$ &     &     &     & 681 &  
 \end{tabular}
\caption{Achieved bandwidth for the 2x3v case on a single GPU using single and double precision. Missing values imply that there is not enough memory on the specific GPU for the specified resolution.}
 \label{tab:gpubandwidth5}
\end{table}
\begin{table}
 \centering
 \begin{tabular}{c|c|c|c|c|c}
     & GTX & TitanV & V100 & \multicolumn{2}{c}{A100} \\ 
 dof &  single & single & single & single & double \\ \hline
 $28^328^3$ & 64 & 281 & 332 & 415 & 503\\
 $28^332^3$ & 66 & 309 & 366 & 462 & 555\\
 $32^332^3$ &    & 355 & 418 & 534 & 590\\
 $32^336^3$ &    &     & 442 & 554 & 626\\
 $36^336^3$ &    &     &     & 460 & 563\\
 $36^340^3$ &    &     &     & 482 &    \\
 $40^340^3$ &    &     &     & 539 &
 \end{tabular}  
\caption{Achieved bandwidth for the 3x3v case on a single GPU using single and double precision. Missing values imply that there is not enough memory on the specific GPU for the specified resolution.}
 \label{tab:gpubandwidth6}
\end{table}

\begin{figure}[h]
\centering
\includegraphics[width=1.0\linewidth]{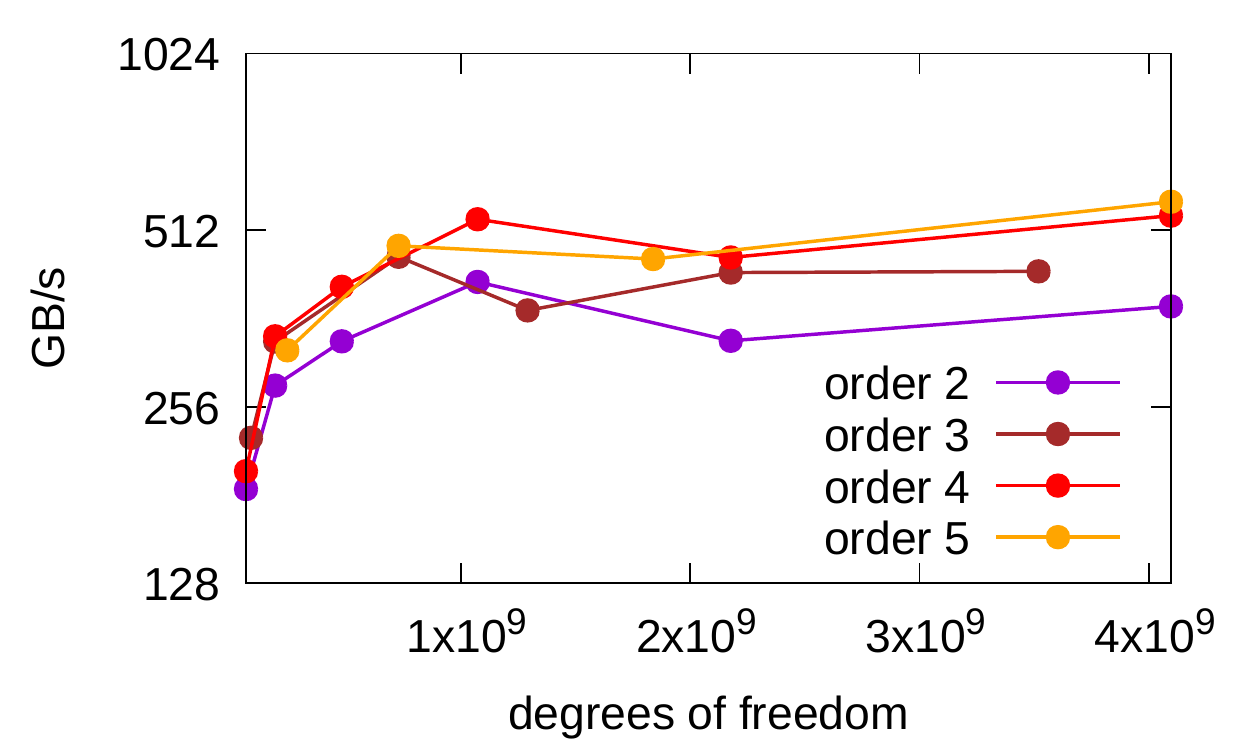}
\caption{\label{fig:scaling6d}Obtained bandwidth in the 3x3v case as a function of the degrees of freedom where different spatial orders of the method are used. The simulations are done in single precision on a A100 GPU.}
\end{figure}

Since single GPU results for the 2x2v and 2x3v case were already presented in \cite{Einkemmer2020GPUs}, we will focus on the six-dimensional 3x3v case in the following. In figure~\ref{fig:scaling6d} a comparison of the achieved bandwidth between different orders of the method in the 3x3v case is shown. The 4\textsuperscript{th} and 5\textsuperscript{th} order method achieve better performance than the 2\textsuperscript{nd} and 3\textsuperscript{rd} order method, which is similar to the behavior observed in \cite{Einkemmer2020GPUs} for the 2x2v and 2x3v case. 

\begin{figure}[h]
\begin{subfigure}{1.0\linewidth}
\centering
 \includegraphics[width=1\textwidth]{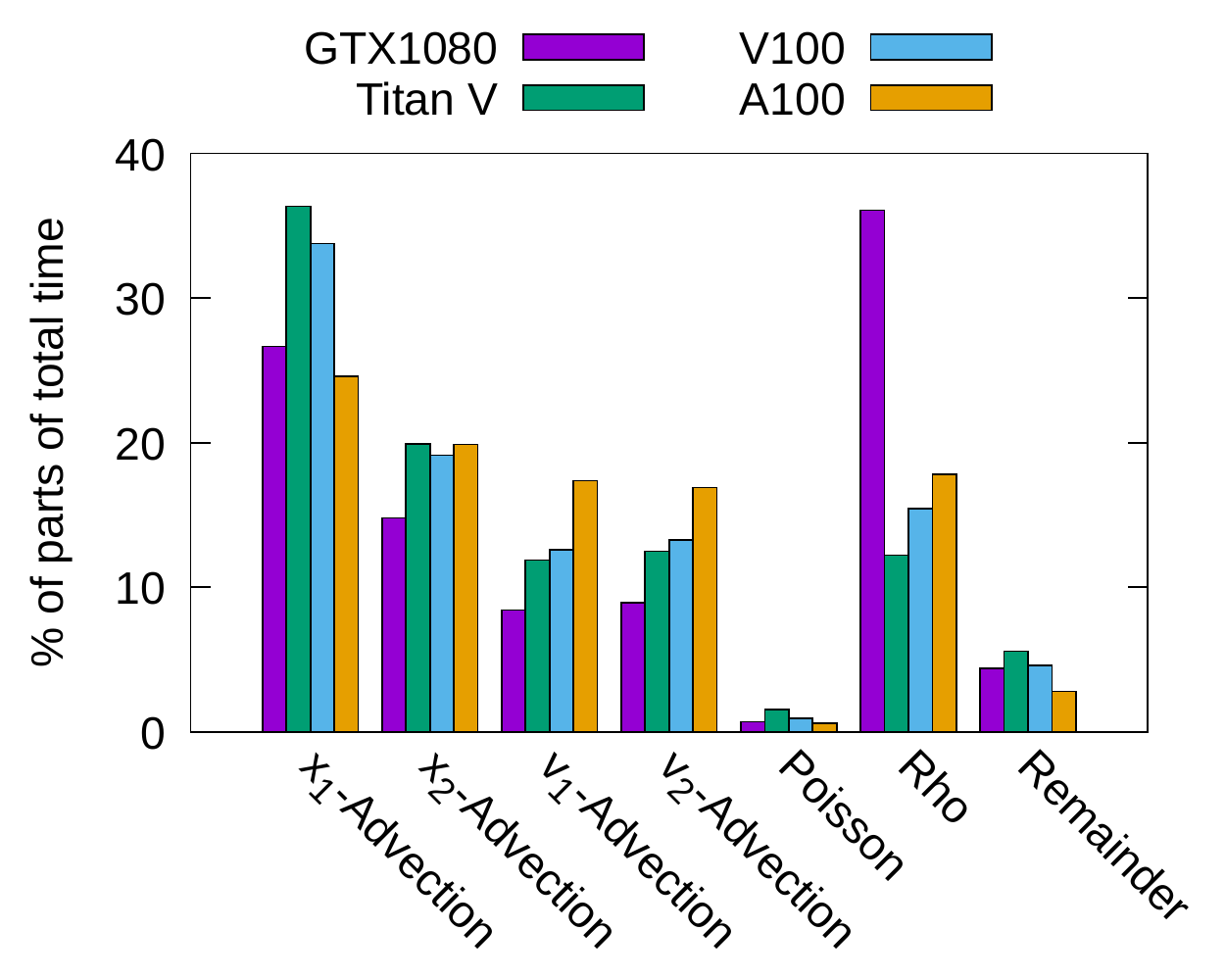}
\end{subfigure}
\begin{subfigure}{1.0\linewidth} 
\centering
 \includegraphics[width=1\textwidth]{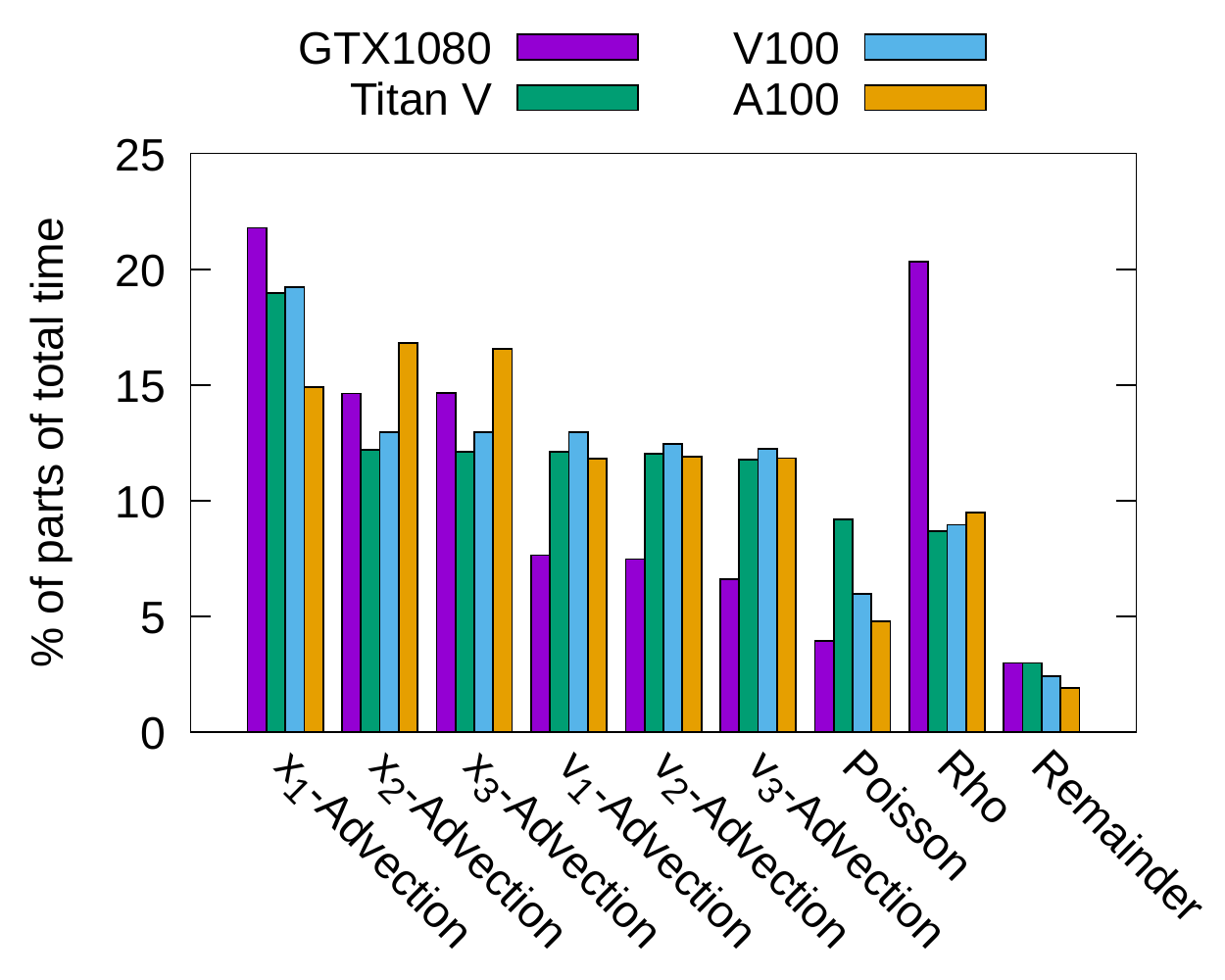}
\end{subfigure}
 \caption{\label{fig:breakdown}Timings for the different parts of the algorithm. A single precision computation with the largest problem that fits into memory for the GPU used has been conducted.}
\end{figure}

In figure~\ref{fig:breakdown} we show the timings for the different parts that constitute the algorithm in the 2x3v and the 3x3v case using different GPUs. The chosen number of degrees of freedom is the maximum amount possible on the specific GPU. In the 2x3v case it can be seen once more that the choice of the Poisson solver has no impact on the overall performance. The main difference between the consumer GPU (the GTX1080) and the enterprise grade GPUs is that the GTX1080 needs more time for the computation of the charge density $\rho$. This comes from the fact that the GTX1080 is able to perform significantly less floating-point operations per second than the other GPUs. Those are important to compute the diagnostic quantities (such as electric energy) in the reduction step. Similar results can be observed for the 3x3v case.

\begin{figure}[h]
  \centering
 \includegraphics[width=0.5\textwidth]{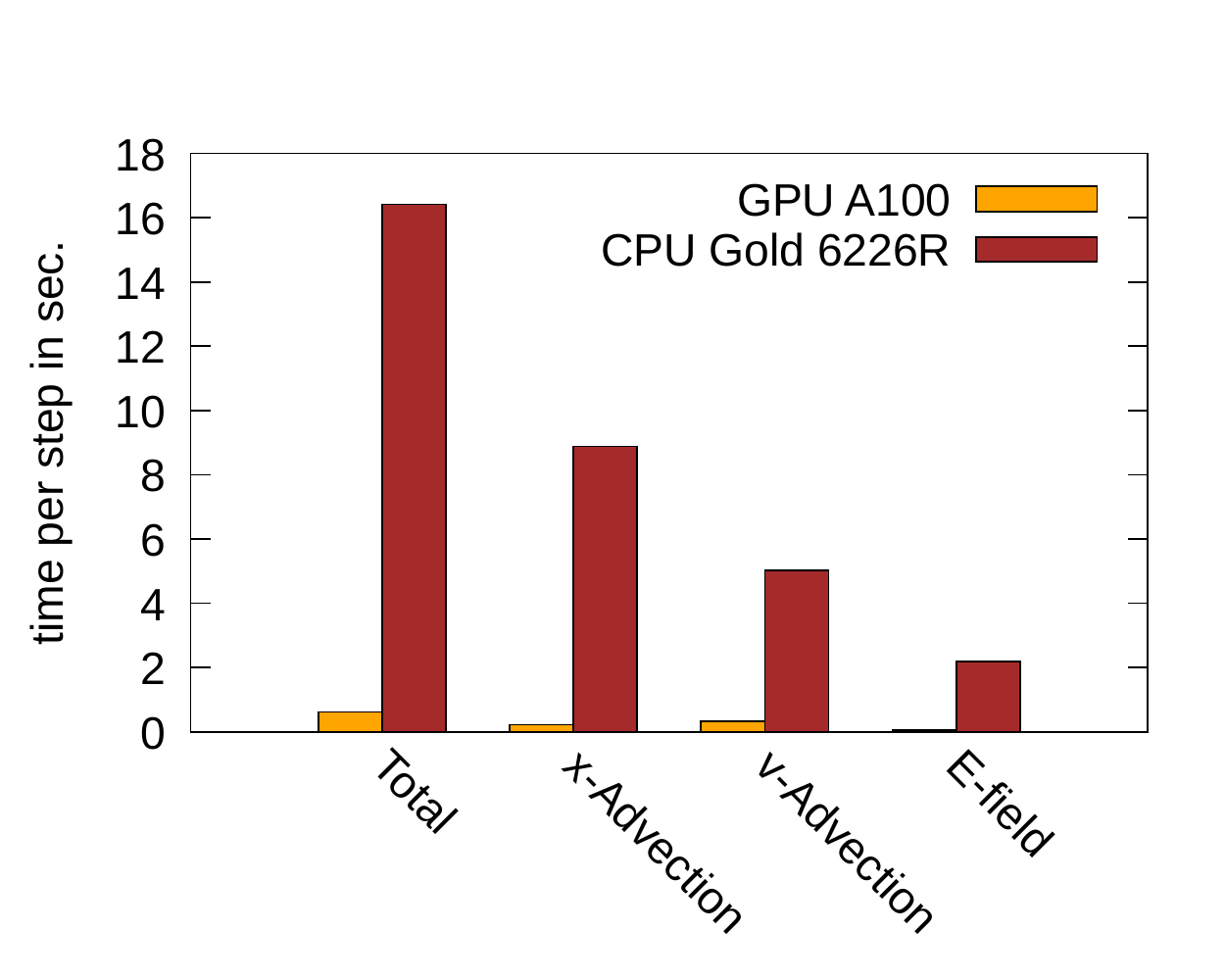}
 \caption{Performance comparison of a CPU based system and a A100 GPU. The GPU is 24.5 times faster than the CPUs. This comparison is obtained using double precision.}
\label{fig:gpuvscpu}
\end{figure}

To conclude this section, we will briefly compare the performance between the GPU implementation (which is the focus of the present work) and the CPU implementation. A comparison for the 3x3v problem between the A100 GPU and a Intel Xeon Gold 6226R dual socket system (having $2\times16$ CPUs) is given in figure~\ref{fig:gpuvscpu}. It can clearly be observed that the GPU drastically outperforms the CPU based system. In fact, every part of the code is computed by the GPU in significantly less time. This, in addition to the four and five dimensional results presented in \cite{Einkemmer2020GPUs}, underscores the advantage of running semi-Lagrangian Vlasov simulations on GPU based systems. We also note that this is enabled by the semi-Lagrangian discontinuous Galerkin method which, in contrast to spline or FFT based interpolation, maps extremely well to the the GPU architecture.

\subsection{Multiple GPU performance}

In this section, we analyze the performance of the code using 4 A100 GPUs on a single node. We increase the number of grid points in two of the velocity dimensions proportionally to the number of GPU used. That is we use weak scaling, which is the only metric of interest for these high-dimensional problems. For the computation we use the Juwels Booster supercomputer. The four A100 GPUs on a single node are connected via NVLink3 to each other. The corresponding performance results are presented in table~\ref{tab:node_performance}, where we also report results for a four A100 node without using NVLink. As we can see, using NVLink results in approximately a factor of $2$ improvement yielding a total single node achieved bandwidth of between $2.05$ and $2.55$ TB/s.

\begin{table}[h]
\centering
 \begin{tabular}{c|c|c|c}
  dim & dof & without NVLink & with NVLink \\ \hline
  2x2v  & $220^2440^2$   & 1821 GB/s & 2437 GB/s  \\
  2x3v  & $72^2144^272$  & 1202 GB/s & 2567 GB/s  \\
  3x3v  & $36^472^2$     & 859 GB/s  & 2052 GB/s  
 \end{tabular}
 \caption{Comparison of the achieved bandwidth using four A100 GPUs on one computation node with and without NVLink. The computations shown here are performed in double precision.}
 \label{tab:node_performance}
\end{table}

\section{Scaling to multiple nodes}

As can be observed in the previous sections, the resulting grid is still relatively coarse. This is especially true in the five and six-dimensional case. In order to run the simulations on finer grids, more nodes are required. In this section, we will report weak scaling results for our code on Juwels Booster. This is the most important metric for practical applications, where a finer grid is often required in order to obtain physically relevant results. The reported measured wall clock time is the average over the computed time steps. We analyze the total computation time and the time for the pure computation of the advections (denoted as Advection in the figures), the time which is required for the data transfer before the advection takes place (denoted as Communication), and the time for computing the electric field (denoted as E field), which includes the computation of the charge density, the time for solving the Poisson problem (including all the required communication) and the broadcast of the electric field. The time step size is chosen such that the CFL-number is close to one. Moreover, the largest possible local grid which fits into GPU memory is chosen. 

\subsubsection*{2x2v}

Two test cases are considered in the 4 dimensional case. First, the full domain is partitioned and distributed over multiple GPUs, and second, only the velocity domain is refined. The latter is quite common as some problems require a higher resolution in velocity space, for example, to shift back in time the (purely numerical) recurrence effect in linear Landau damping simulations \cite{EinkemmerRecurrence}. In table~\ref{tab:2x2v} the considered grids are listed.

\begin{table}[h]
\begin{subtable}{0.5\textwidth}
\centering
 \begin{tabular}{r|c|c|c|c}
  \multicolumn{5}{c}{local grid 2x2v} \\ \hline
 dim & $x_1$ & $x_2$ & $v_1$ & $v_2$ \\ 
 dof & 220 & 220 & 220 & 220 \vspace{0.2cm} \\ 
 \end{tabular}
 \end{subtable}

  \vspace{0.2cm}
 \begin{subtable}{0.5\textwidth}
 \centering
 \begin{tabular}{r|c|c|c|c}
  \#GPUs & \multicolumn{4}{|c}{decomp. in $xv$} \\ \hline
   1   & 1 & 1 & 1 & 1  \\ 
   4   & 1 & 1 & 2 & 2  \\ 
   8   & 1 & 2 & 2 & 2  \\
   16  & 2 & 2 & 2 & 2  \\
   32  & 2 & 2 & 2 & 4  \\
   64  & 2 & 2 & 4 & 4  \\
   128 & 2 & 4 & 4 & 4  \\
   256 & 4 & 4 & 4 & 4  \\
   512 & 4 & 4 & 4 & 8  \\
  1024 & 4 & 4 & 8 & 8  \\
 \end{tabular}
 \end{subtable}

  \vspace{0.2cm}
 \begin{subtable}{0.5\textwidth}
  \centering
  \begin{tabular}{r|c|c|c|c}
  \#GPUs & \multicolumn{4}{|c}{decomp. in $v$} \\ \hline
   1    & 1 & 1 & 1 & 1  \\ 
   4    & 1 & 1 & 2 & 2  \\ 
   16   & 1 & 1 & 4 & 4  \\
   64   & 1 & 1 & 8 & 8  \\
   144  & 1 & 1 & 12 & 12  \\
   256  & 1 & 1 & 16 & 16  \\
   484  & 1 & 1 & 22 & 22  \\
   1024 & 1 & 1 & 32 & 32  \\
  \end{tabular}
 \end{subtable}
\caption{For the 2x2v problem we use a local grid of $220^4$ degrees of freedom on each GPU. In the multi-node setting the GPUs are either equally divided among the different coordinate axis (middle table) or only the velocity direction is refined (bottom table).}
\label{tab:2x2v}
\end{table}

\begin{figure}[h]
 \centering
 \includegraphics[width=0.5\textwidth]{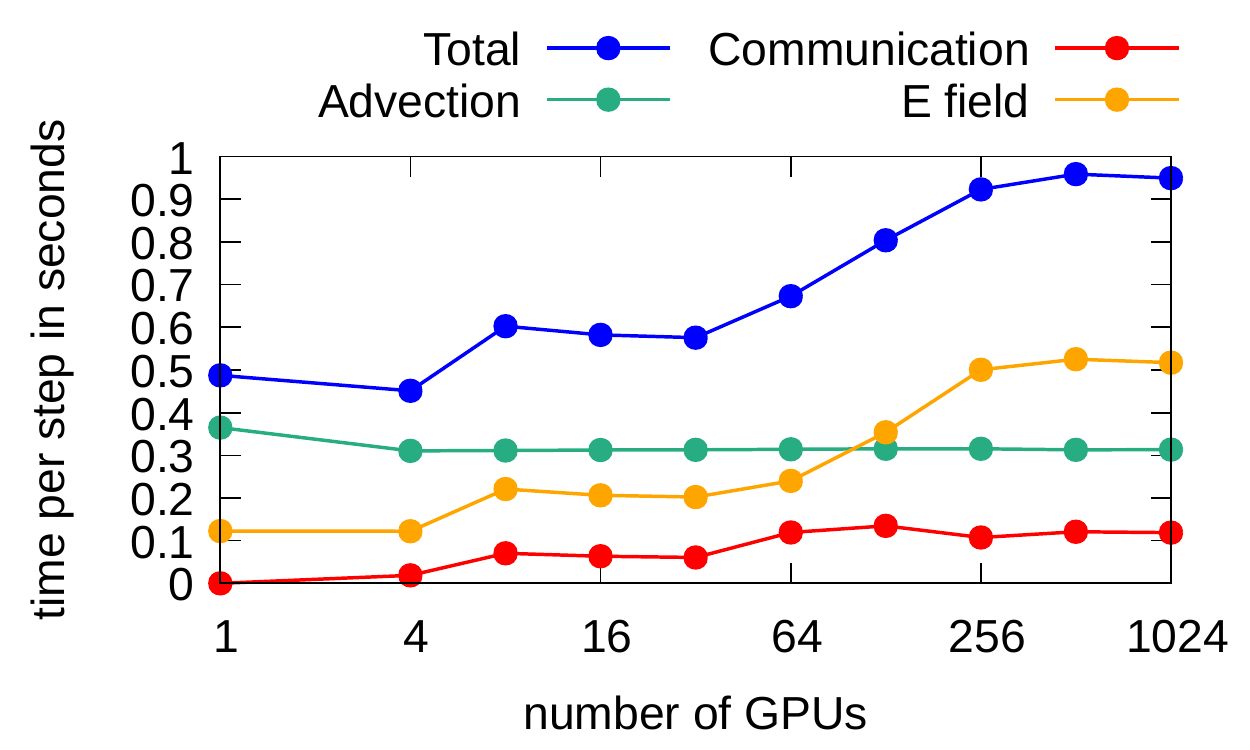}
 \caption{Weak scaling in the 2x2v dimensional setting using multiple GPUs for each dimension. A time step size of 0.015 and the CGdG method to solve the Poisson problem is used.}
 \label{fig:scaling4d}
\end{figure}

Let us first consider the case when the full domain is distributed over multiple GPUs. The corresponding results are shown in figure \ref{fig:scaling4d}. A time step of 0.015 is chosen in order to keep the CFL number below 1, and the CGdG method is used to solve the Poisson problem. The largest grid here has $880^21760^2$ degrees of freedom. It can be observed that the time to perform the advection remains constant. This holds for each configuration in this chapter. The communication time in this setting is relatively small as at each time step only 2 GB of data has to be transferred to neighboring MPI processes. The time for computing the electric field increases from 4 to 16 GPUs since a larger problem has to be solved, i.e., going from a grid size of $220^2$ to $440^2$ and thus more iterations are required to solve the linear system. The system size stays then constant until 64 GPUs are used. Subsequently, the problem size in physical space increases again. That is, a grid size in the space direction of $880^2$ is used when 256  or more GPUs are used. This again requires more iterations and the wall clock time increases again. The total time per step rises from about 0.5 seconds to around 1.0 seconds. Thus running the simulation on $1024$ GPUs a parallel efficiency of approximately 50\% is achieved.

\begin{figure}[h]
 \centering
 \includegraphics[width=0.5\textwidth]{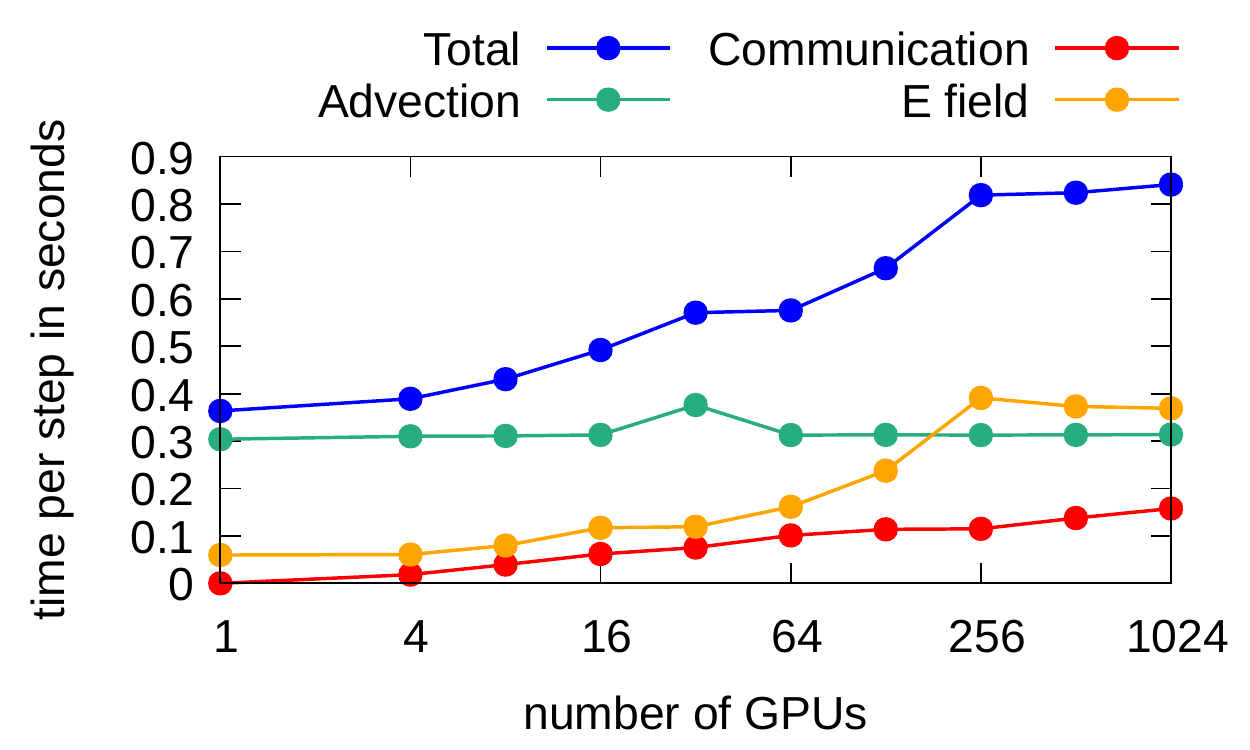}
 \caption{Weak scaling in the 2x2v dimensional setting using multiple GPUs for each dimension. A time step case size of 0.015 and a FFT based solver for the Poisson problem is used.}
 \label{fig:scaling4dfft}
\end{figure}

Since the impact of the Poisson solver on the overall performance, due to the large number of required iterations in the 2x2v setting is quite high, we also investigate the use of a FFT based method to solve the Poisson problem. The results are shown in figure \ref{fig:scaling4dfft}. The FFT based solver is faster on a single node but scaling is a challenge. Thus, the overall parallel efficiency decreases a little, while overall the run time on 1024 GPUs is still reduced by approximately 10\% compared to the CGdG method.

\begin{figure}[h]
 \centering
 \includegraphics[width=0.5\textwidth]{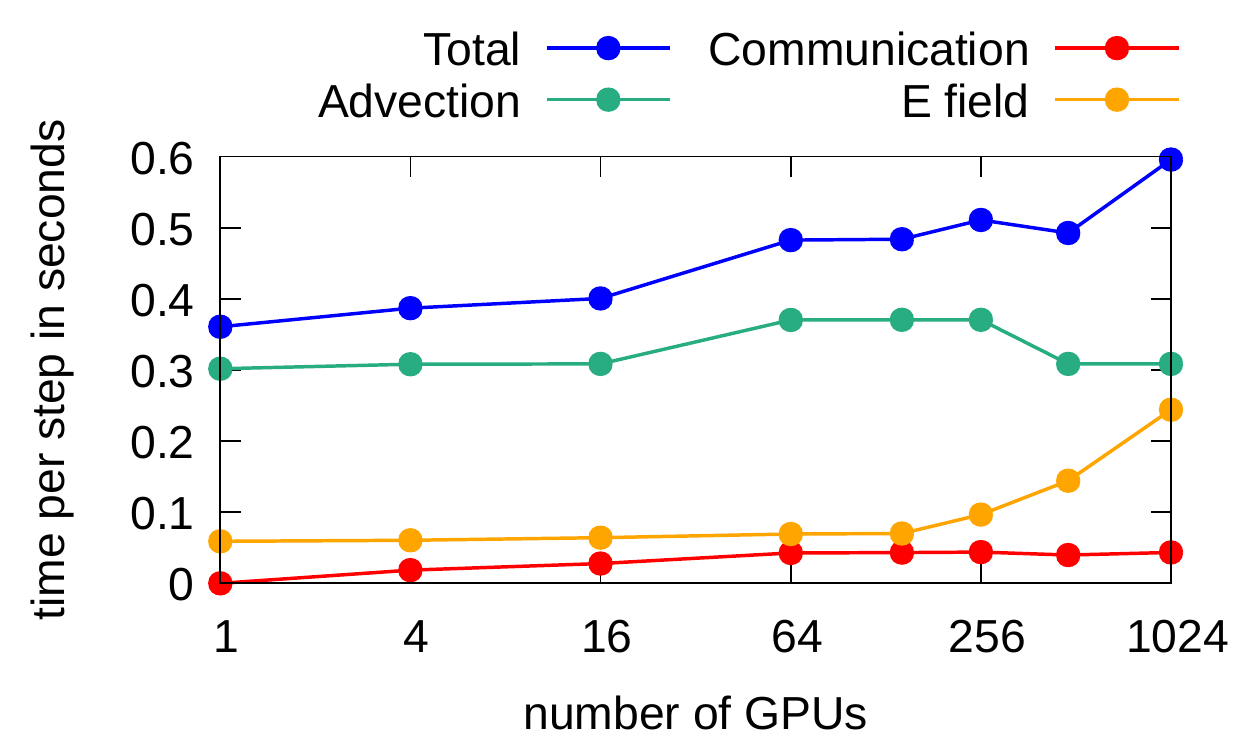}
 \caption{Weak scaling in the 2x2v dimensional setting using multiple GPUs to increase the degrees of freedom in the velocity dimensions only.}
 \label{fig:scaling4dv}
\end{figure}

Next, we analyze the behavior when the degrees of freedom are increased in the velocity directions only. Here, since in physical space no parallel numerical method for the computations is required, the FFT based Poisson solver is used since it is faster. In figure \ref{fig:scaling4dv} the weak scaling results are given. It can be observed that the required time for computing the electric field increases when more than 144 GPUs are used. Here, in contrast to the results shown in figure~\ref{fig:scaling4d} and~\ref{fig:scaling4dfft}, not the time for solving the Poisson problem increases, but the required time to compute the charge density, i.e., the call to \texttt{MPI\_Reduce}. The reason for this is that now a very fine resolution of $7040^2$ grid points in the velocity directions is used. The total run time increases from about $0.36$ to $0.6$, thus a parallel efficiency of 60\% is achieved on $1024$ GPUs.

\begin{figure}[h]
 \centering
 \includegraphics[width=0.5\textwidth]{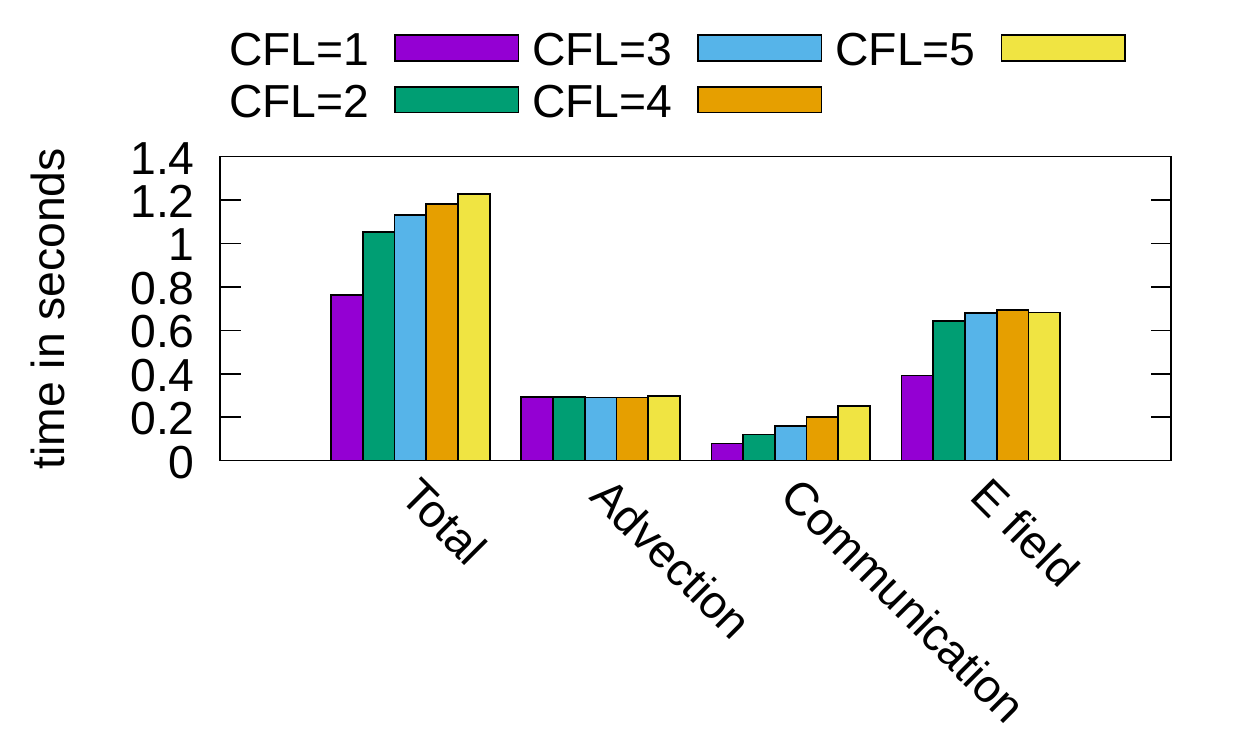}
 \caption{Timings of the different parts of the algorithm using CFL numbers larger than one. A local grid of $216^4$ points is used and the simulation is run on 256 GPUs. We increase the time step from $0.015$ (CFL=1) up to $0.095$ (CFL=5).}
 \label{fig:cfl4d}
\end{figure}

The simulations presented previously were done with a relatively small time step size of $0.015$ in order to keep the CFL number in the spatial directions below one. However, the semi-Lagrangian discontinuous Galerkin scheme is stable also for larger time steps. We will thus investigate the parallel efficiency for a CFL number larger than one. The main additional issue here is that the amount of boundary data that need to be communicated to neighboring processes increases linearly with the CFL number. We use a local grid with a resolution of $216^4$ points and in total $256$ GPUs, thus distributing the domain in each direction over 4 GPUs. The local grid here is slightly smaller than in the previous simulations since additional memory is required to store the boundary data. The CFL number increases from 1 to 5 by increasing the time step size by $0.02$ until we reach a step size of $0.095$ (where the CFL number becomes 5). The results are reported in figure~\ref{fig:cfl4d}. It can be observed that the time for communication increases continuously, which is expected since more boundary cells have to be send to the neighboring processes. Furthermore, the time for computing the advection remains almost constant, and the time for computing the electric field increases. This comes from the fact that when solving the Poisson problem iteratively, the starting point is the electric potential obtained at the previous time step. Thus, when a larger time step is used, the initial estimate used in the iterative Poisson solver is farther away from the solution and thus more iterations are required for convergence. Nevertheless, we note that the run for a CFL number smaller than $1$ to a CFL number of $5$ increases by approximately 50\%. Thus, running the numerical simulation at a CFL number of $5$ still results in a speedup of approximately a factor $3.3$ (despite the increased cost of communication per time step).

\subsubsection*{2x3v}

\begin{figure}[h]
 \centering
 \includegraphics[width=0.5\textwidth]{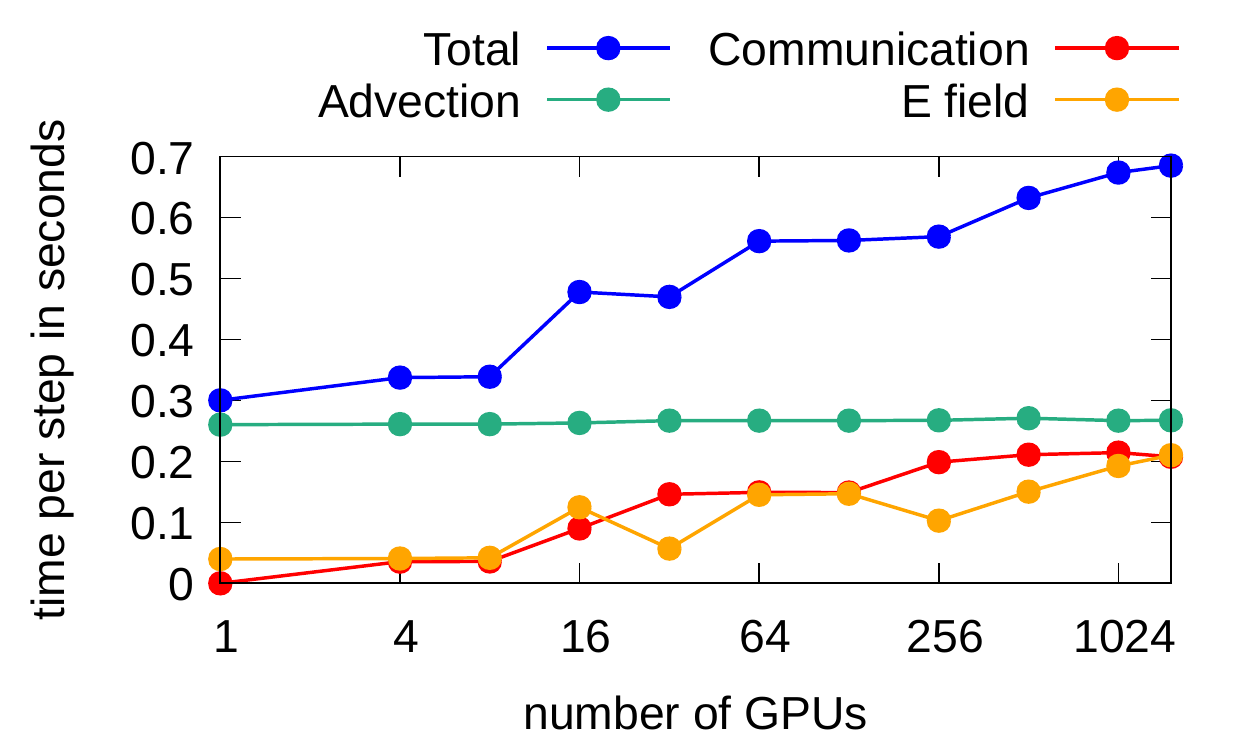}
 \caption{Weak scaling in the 2x3v case using multiple GPUs for each dimension. A time step size of $0.05$ and the CGdG method to solve the Poisson problem is used.}
 \label{fig:scaling5d}
\end{figure}

\begin{figure}[h]
 \centering
 \includegraphics[width=0.5\textwidth]{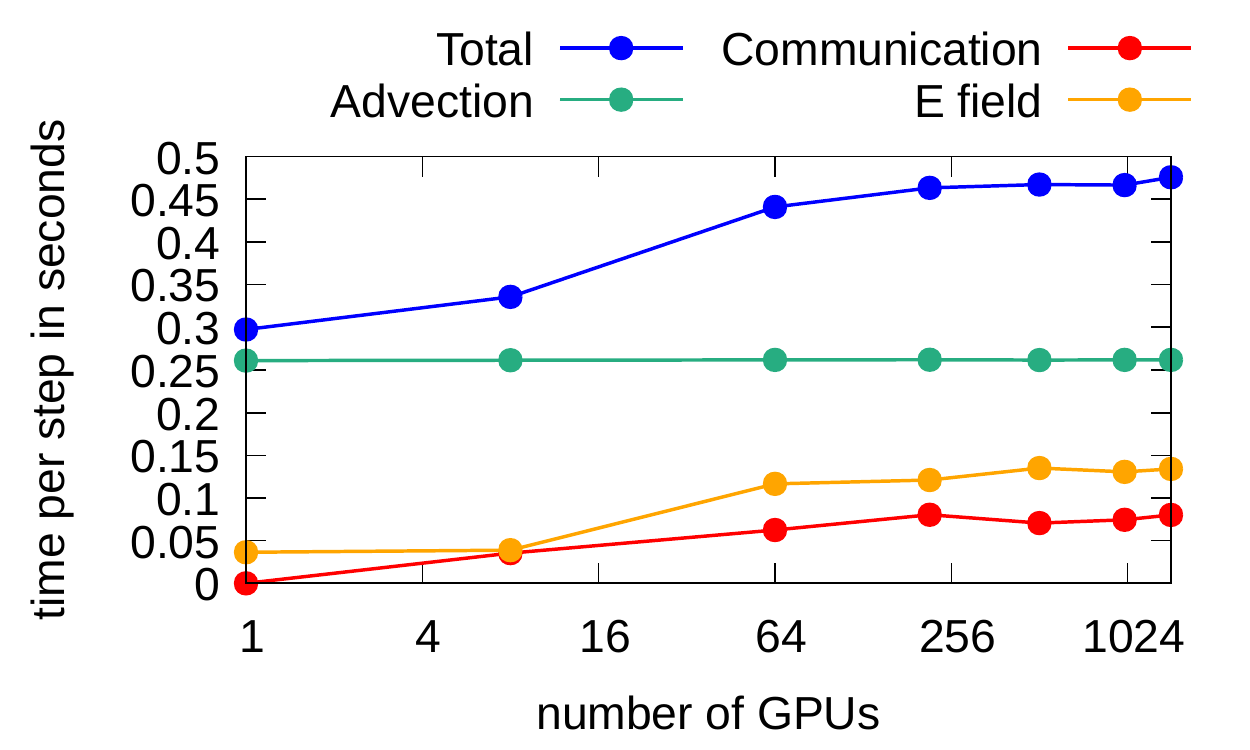}
 \caption{Weak scaling in the 2x3v dimensional setting using multiple GPUs for velocity dimensions only.}
 \label{fig:scaling5dv}
\end{figure}

In figure \ref{fig:scaling5d} and \ref{fig:scaling5dv} the scaling results for the 2x3v case are shown. Again, as in the 2x2v case, two configurations are considered, namely the full grid refinement and increasing the degrees of freedom in the velocity domain only. The considered configurations are listed in table~\ref{tab:2x3v}.

\begin{table}[h]
 \begin{subtable}{0.5\textwidth}
 \centering
 \begin{tabular}{r|c|c|c|c|c}
  \multicolumn{6}{c}{local grid $2x3v$} \\ \hline
 dim & $x_1$ & $x_2$ & $v_1$ & $v_2$ & $v_3$ \\ 
 dof & 72 & 72 & 72 & 72 & 72 \\ 
 \end{tabular}
 \end{subtable}

 \vspace{0.2cm}
 \begin{subtable}{0.5\textwidth}
 \centering
 \begin{tabular}{r|c|c|c|c|c}
  \#GPUs & \multicolumn{5}{|c}{decomp. in $xv$} \\ \hline
   1   & 1 & 1 & 1 & 1 & 1  \\ 
   4   & 1 & 1 & 2 & 2 & 1  \\ 
   8   & 1 & 1 & 2 & 2 & 2  \\
   16  & 1 & 2 & 2 & 2 & 2  \\
   32  & 2 & 2 & 2 & 2 & 2  \\
   64  & 2 & 2 & 2 & 2 & 4  \\
   128 & 2 & 2 & 2 & 4 & 4  \\
   256 & 2 & 2 & 4 & 4 & 4  \\
   512 & 2 & 4 & 4 & 4 & 4  \\
  1024 & 4 & 4 & 4 & 4 & 4  \\
  1536 & 4 & 4 & 4 & 4 & 6 
 \end{tabular}
 \end{subtable}

 \vspace{0.2cm}
 \begin{subtable}{0.5\textwidth}
 \centering
 \begin{tabular}{r|c|c|c|c|c}
  \#GPUs & \multicolumn{5}{|c}{decomp. in $v$} \\ \hline
   1   & 1 & 1 & 1 & 1 & 1  \\ 
   8   & 1 & 1 & 2 & 2 & 2  \\
   64  & 1 & 1 & 4 & 4 & 4  \\
   216 & 1 & 1 & 6 & 6 & 6  \\
   512 & 1 & 1 & 8 & 8 & 8  \\
  1000 & 1 & 1 & 10 & 10 & 10 \\
  1440 & 1 & 1 & 12 & 12 & 10 
 \end{tabular}
 \end{subtable}
\caption{For the 2x3v problem we use a local grid of $72^5$ degrees of freedom on each GPU. In the multi-node setting the GPUs are either equally divided among the different coordinate axis (middle table) or only the velocity direction is refined (bottom table).}
\label{tab:2x3v}
\end{table}

When the full domain is distributed over multiple GPUs, a time step of 0.05 is chosen which implies a CFL number below 1. The CGdG method is used to solve the Poisson problem. The largest problem has a grid of $288^4432$ and is run on 1536 GPUs. In figure~\ref{fig:scaling5d} it can be observed that the computation of the electric field has much less impact on the overall performance than in the 2x2v setting. This is mainly because solving the Poisson problem with the CGdG requires less iterations due the smaller number of degrees of freedom per direction. Nevertheless, also here, as expected, an increase of required wall time can be observed if the problem size increases. The time for the communication which is required to perform the advection has more impact on the overall performance than in the 2x2v setting. The CGdG method allows us to limit the impact to performance and is the preferred choice for this five-dimensional problem. For 1536 GPUs we achieve a parallel efficiency of 43\%.

In problems where only additional degrees of freedom are added in the velocity directions, similar results as have been obtained in the 2x2v case can be observed. In this case the FFT based solver is the preferred choice. The corresponding weak scaling results are shown in figure~\ref{fig:scaling5dv}, where we observe almost ideal scaling for up to 1440 GPUs, The parallel efficiency in this case is approximately 67\%.

\subsubsection*{3x3v}

In the six dimensional setting the only reasonable choice is to distribute the problem in each direction (if a distribution is done only in the velocity direction the number of degrees of freedom in physical space is too small to be of practical interest). The used configurations are listed in table~\ref{tab:3x3v}. Using 1536 GPUs, results in a grid of size $72^1108^1144^4$ in double precision, which is still relatively coarse, and $80^1 120^1 160^4$ in single precision. 
\begin{table}[h]
\centering
 \begin{tabular}{r|c|c|c|c|c|c}
  \multicolumn{7}{c}{local grid 3x3v double} \\ \hline
 dim & $x_1$ & $x_2$ & $x_3$ & $v_1$ & $v_2$ & $v_3$ \\ 
 dof & 36  &  36 & 36 &  36 &  36 & 36\vspace{0.2cm} \\ 
 \multicolumn{7}{c}{local grid 3x3v single} \\ \hline
 dim & $x_1$ & $x_2$ & $x_3$ & $v_1$ & $v_2$ & $v_3$ \\ 
 dof & 40 & 40 & 40 & 40 & 40 & 40\vspace{0.2cm} \\ 
  \#GPUs & \multicolumn{6}{|c}{decomp. in $xv$} \\ \hline
   1   & 1 & 1 & 1 & 1 & 1 & 1  \\ 
   4   & 1 & 1 & 1 & 1 & 2 & 2  \\ 
   8   & 1 & 1 & 1 & 2 & 2 & 2  \\
   16  & 1 & 1 & 2 & 2 & 2 & 2  \\
   32  & 1 & 2 & 2 & 2 & 2 & 2  \\
   64  & 2 & 2 & 2 & 2 & 2 & 2  \\
   128 & 2 & 2 & 2 & 2 & 2 & 4  \\
   256 & 2 & 2 & 2 & 2 & 4 & 4  \\
   512 & 2 & 2 & 2 & 4 & 4 & 4  \\
  1024 & 2 & 2 & 4 & 4 & 4 & 4  \\
  1536 & 2 & 3 & 4 & 4 & 4 & 4 
 \end{tabular}
\caption{Configurations in the 3x3v case using multiple GPUs.}
\label{tab:3x3v}
\end{table}

\begin{figure}[h]
 \centering
 \includegraphics[width=0.5\textwidth]{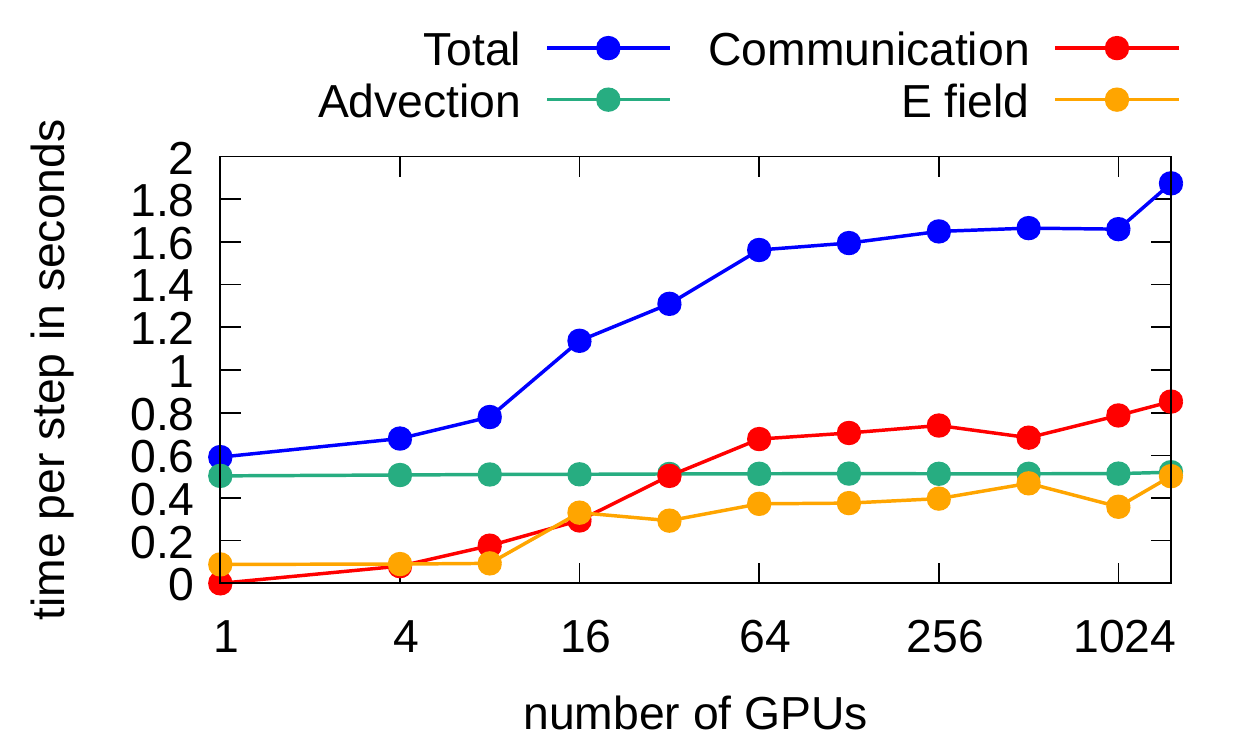}
 \caption{Weak scaling in the 3x3v case using multiple GPUs. The spatial configuration is listed in table~\ref{tab:3x3v}. Computations are done in double precision.}
 \label{fig:scaling6dx}
\end{figure}

In figure \ref{fig:scaling6dx} the scaling results for the 3x3v case are shown. Due to the coarse grid, a time step of 0.1 is chosen which is enough to keep the CFL number below 1. The CGdG solver is the preferred choice to solve the Poisson problem. It can be observed that the communication time increases significantly until 64 GPUs are used and remains almost constant afterwards. The reason for this behavior is that for a six-dimensional problem 64 GPUs are required so that in each direction we use more than a single GPU. Due to the nearest neighbor exchange any further parallelization does, in principle, not contribute to the overall communication time. The time per step increases from 0.6 seconds using one GPU to 1.9 seconds using 1536 GPUs, thus a parallel efficiency of approximately 30\% is achieved.

\begin{figure}[h]
 \centering
 \includegraphics[width=0.5\textwidth]{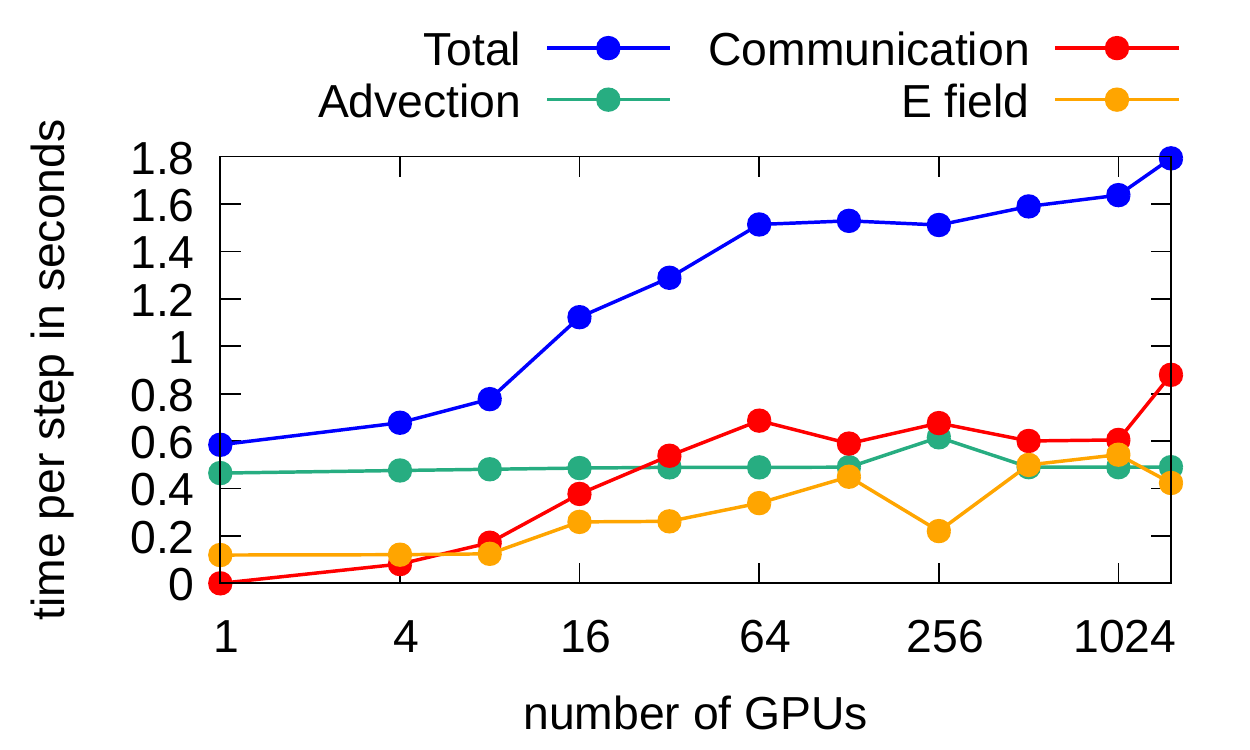}
 \caption{Weak scaling in the $3x3v$ case using multiple GPUs. The configuration is listed in table~\ref{tab:3x3v}. Computations are performed in single precision.}
 \label{fig:scaling6dsingle}
\end{figure}

In figure~\ref{fig:scaling6dsingle} the results using single precision are shown. Since less memory is required, approximately 1.9 times more degrees of freedom can be considered compared to double precision. We observe a similar parallel efficiency of approximately 33\%.

\section{Conclusion}
\label{sec:conclusion}

In this work we have demonstrated that semi-Lagrangian discontinuous Galerkin schemes can be used to run five and six-dimensional Vlasov simulations on large scale GPU equipped supercomputers. The local nature of this method gives excellent performance on a single node using 4 GPUs (between $2$ and $2.5$ TB/s depending on the configuration) and scales well to up to 1500 GPUs on Juwels Booster with a parallel efficiency of between 30\% and 67\% (depending on the configuration). Since GPU based supercomputers will play an important role in exascale systems and, most likely, in the future of high-performance computing, the results demonstrated in this work pave the way towards running large scale Vlasov simulations on such systems.

\section*{Acknowledgment}
We acknowledge PRACE for awarding us access to JUWELS at GCS@FZJ, Germany. The computational results presented have been achieved in part using the Vienna Scientific Cluster (VSC).

\section*{Funding}
The author(s) disclosed receipt of following financial support for the research, authorship, and/or publication of this article: This project has received funding from the European Union's Horizon 2020 research and innovation programme under the Marie Skłodowska-Curie grant agreement No 847476. The views and opinions expressed herein do not necessarily reflect those of the European Commission.

\section*{Declaration of Conflicting Interests}
The author(s) declared no potential conflicts of interest with respect to the research, authorship, and/or publication of this article.

\bibliography{literatur}

\begin{thebibliography}{32}
\providecommand{\natexlab}[1]{#1}
\providecommand{\url}[1]{\texttt{#1}}
\providecommand{\urlprefix}{URL }
\expandafter\ifx\csname urlstyle\endcsname\relax
  \providecommand{\doi}[1]{DOI:\discretionary{}{}{}#1}\else
  \providecommand{\doi}{DOI:\discretionary{}{}{}\begingroup
  \urlstyle{rm}\Url}\fi

\bibitem[{Bigot et~al.(2013)Bigot, Grandgirard, Latu, Passeron, Rozar and
  Thomine}]{bigot2013scaling}
Bigot J, Grandgirard V, Latu G, Passeron C, Rozar F and Thomine O (2013)
  {Scaling Gysela code beyond 32K-cores on BlueGene/Q}.
\newblock In: \emph{ESAIM: Proceedings}, volume~43. EDP Sciences, pp. 117--135.

\bibitem[{Camporeale et~al.(2016)Camporeale, Delzanno, Bergen and
  Moulton}]{camporeale2016velocity}
Camporeale E, Delzanno G, Bergen B and Moulton J (2016) {On the velocity space
  discretization for the Vlasov--Poisson system: Comparison between implicit
  Hermite spectral and Particle-in-Cell methods}.
\newblock \emph{Computer Physics Communications} 198: 47--58.

\bibitem[{Casas et~al.(2017)Casas, Crouseilles, Faou and
  Mehrenberger}]{casas2017high}
Casas F, Crouseilles N, Faou E and Mehrenberger M (2017) {High-order
  Hamiltonian splitting for the Vlasov--Poisson equations}.
\newblock \emph{Numerische Mathematik} 135(3): 769--801.

\bibitem[{Cheng and Knorr(1976)}]{CHENG1976}
Cheng C and Knorr G (1976) {The integration of the Vlasov equation in
  configuration space}.
\newblock \emph{Journal of Computational Physics} 22(3): 330--351.

\bibitem[{Crouseilles et~al.(2015)Crouseilles, Einkemmer and
  Faou}]{crouseilles2015hamiltonian}
Crouseilles N, Einkemmer L and Faou E (2015) {Hamiltonian splitting for the
  Vlasov--Maxwell equations}.
\newblock \emph{Journal of Computational Physics} 283: 224--240.

\bibitem[{Crouseilles et~al.(2009)Crouseilles, Latu and
  Sonnendr{\"u}cker}]{crouseilles2009parallel}
Crouseilles N, Latu G and Sonnendr{\"u}cker E (2009) {A parallel Vlasov solver
  based on local cubic spline interpolation on patches}.
\newblock \emph{Journal of Computational Physics} 228(5): 1429--1446.

\bibitem[{Crouseilles et~al.(2011)Crouseilles, Mehrenberger and
  Vecil}]{crouseilles2011discontinuous}
Crouseilles N, Mehrenberger M and Vecil F (2011) {Discontinuous Galerkin
  semi-Lagrangian method for Vlasov-Poisson}.
\newblock In: \emph{ESAIM: Proceedings}, volume~32. EDP Sciences, pp. 211--230.

\bibitem[{Einkemmer(2017)}]{einkemmer2017study}
Einkemmer L (2017) {A study on conserving invariants of the Vlasov equation in
  semi-Lagrangian computer simulations}.
\newblock \emph{Journal of Plasma Physics} 83(2).

\bibitem[{Einkemmer(2019)}]{Einkemmer20194d}
Einkemmer L (2019) {A performance comparison of semi-Lagrangian discontinuous
  Galerkin and spline based Vlasov solvers in four dimensions}.
\newblock \emph{Journal of Computational Physics} 376: 937--951.

\bibitem[{Einkemmer(2020)}]{Einkemmer2020GPUs}
Einkemmer L (2020) {Semi-Lagrangian Vlasov simulation on GPUs}.
\newblock \emph{Computer Physics Communications} 254: 107351.

\bibitem[{Einkemmer and Joseph(2021)}]{Einkemmer2021lr}
Einkemmer L and Joseph I (2021) {A mass, momentum, and energy conservative
  dynamical low-rank scheme for the Vlasov equation}.
\newblock \emph{Journal of Computational Physics} : 110495.

\bibitem[{Einkemmer and Lubich(2018)}]{einkemmer2018low}
Einkemmer L and Lubich C (2018) A low-rank projector-splitting integrator for
  the {V}lasov--{P}oisson equation.
\newblock \emph{SIAM Journal on Scientific Computing} 40(5): B1330--B1360.

\bibitem[{Einkemmer and Ostermann(2014{\natexlab{a}})}]{EinkemmerRecurrence}
Einkemmer L and Ostermann A (2014{\natexlab{a}}) {A strategy to suppress
  recurrence in grid-based Vlasov solvers}.
\newblock \emph{{The European Physical Journal D}} 68: 197.

\bibitem[{Einkemmer and Ostermann(2014{\natexlab{b}})}]{einkemmer2014b}
Einkemmer L and Ostermann A (2014{\natexlab{b}}) {Convergence analysis of a
  discontinuous Galerkin/Strang splitting approximation for the Vlasov--Poisson
  equations}.
\newblock \emph{SIAM Journal on Numerical Analysis} 52(2): 757--778.

\bibitem[{Einkemmer and Ostermann(2014{\natexlab{c}})}]{einkemmer2014a}
Einkemmer L and Ostermann A (2014{\natexlab{c}}) {Convergence analysis of
  Strang splitting for Vlasov-type equations}.
\newblock \emph{SIAM Journal on Numerical Analysis} 52(1): 140--155.

\bibitem[{Epshteyn and Rivière(2007)}]{EPSHTEYN2007}
Epshteyn Y and Rivière B (2007) {Estimation of penalty parameters for
  symmetric interior penalty Galerkin methods}.
\newblock \emph{Journal of Computational and Applied Mathematics} 206(2):
  843--872.

\bibitem[{Filbet et~al.(2001)Filbet, Sonnendr{\"u}cker and
  Bertrand}]{filbet2001conservative}
Filbet F, Sonnendr{\"u}cker E and Bertrand P (2001) {Conservative numerical
  schemes for the Vlasov equation}.
\newblock \emph{Journal of Computational Physics} 172(1): 166--187.

\bibitem[{Filbet and Sonnendrücker(2003)}]{Filbet2003interpolations}
Filbet F and Sonnendrücker E (2003) {Comparison of Eulerian Vlasov solvers}.
\newblock \emph{Computer Physics Communications} 150(3): 247--266.

\bibitem[{Guo and Cheng(2016)}]{sparseGrid2016}
Guo W and Cheng Y (2016) {A sparse grid discontinuous Galerkin method for
  high-dimensional transport equations and its application to kinetic
  simulations}.
\newblock \emph{SIAM Journal on Scientific Computing} 38(6): A3381--A3409.

\bibitem[{Hariri et~al.(2016)Hariri, Tran, Jocksch, Lanti, Progsch, Messmer,
  Brunner, Gheller and Villard}]{Hariri2016pic}
Hariri F, Tran T, Jocksch A, Lanti E, Progsch J, Messmer P, Brunner S, Gheller
  C and Villard L (2016) {A portable platform for accelerated PIC codes and its
  application to GPUs using OpenACC}.
\newblock \emph{Computer Physics Communications} 207: 69--82.

\bibitem[{Howes et~al.(2008)Howes, Dorland, Cowley, Hammett, Quataert,
  Schekochihin and Tatsuno}]{howes2008}
Howes G, Dorland W, Cowley S, Hammett G, Quataert E, Schekochihin A and Tatsuno
  T (2008) {Kinetic simulations of magnetized turbulence in astrophysical
  plasmas}.
\newblock \emph{Physical Review Letters} 100: 065004.

\bibitem[{Klimas and Farrell(1994)}]{klimas1994}
Klimas A and Farrell W (1994) {A splitting algorithm for Vlasov simulation with
  filamentation filtration}.
\newblock \emph{Journal of Computational Physics} 110(1): 150--163.

\bibitem[{Klimas and Vi{\~n}as(2018)}]{klimas2018}
Klimas A and Vi{\~n}as A (2018) {Absence of recurrence in Fourier--Fourier
  transformed Vlasov--Poisson simulations}.
\newblock \emph{Journal of Plasma Physics} 84(4).

\bibitem[{Kormann et~al.(2019)Kormann, Reuter and Rampp}]{Kormann2019}
Kormann K, Reuter K and Rampp M (2019) {A massively parallel semi-Lagrangian
  solver for the six-dimensional Vlasov–Poisson equation}.
\newblock \emph{The International Journal of High Performance Computing
  Applications} 33(5): 924--947.

\bibitem[{Kormann and Sonnendr{\"u}cker(2016)}]{kormann2016}
Kormann K and Sonnendr{\"u}cker E (2016) Sparse {G}rids for the
  {V}lasov--{P}oisson {E}quation.
\newblock In: \emph{Sparse Grids and Applications - Stuttgart 2014}. Springer
  International Publishing, pp. 163--190.

\bibitem[{Mehrenberger et~al.(2013)Mehrenberger, Steiner, Marradi, Crouseilles,
  Sonnendr{\"u}cker and Afeyan}]{mehrenberger2013vlasov}
Mehrenberger M, Steiner C, Marradi L, Crouseilles N, Sonnendr{\"u}cker E and
  Afeyan B (2013) {Vlasov on GPU (VOG project)}.
\newblock In: \emph{ESAIM: Proceedings}, volume~43. pp. 37--58.

\bibitem[{NVIDIA(2018)}]{CUB}
NVIDIA (2018) Cub.
\newblock https://nvlabs.github.io/cub/.

\bibitem[{Rivière(2008)}]{dgbook}
Rivière B (2008) \emph{Discontinuous Galerkin Methods for Solving Elliptic and
  Parabolic Equations: Theory and Implementation}.
\newblock Society for Industrial and Applied Mathematics.
\newblock \doi{10.1137/1.9780898717440}.

\bibitem[{Rossmanith and Seal(2011)}]{rossmanith2011}
Rossmanith J and Seal D (2011) {A positivity-preserving high-order
  semi-Lagrangian discontinuous Galerkin scheme for the Vlasov--Poisson
  equations}.
\newblock \emph{Journal of Computational Physics} 230(16): 6203--6232.

\bibitem[{Sonnendr{\"u}cker et~al.(1999)Sonnendr{\"u}cker, Roche, Bertrand and
  Ghizzo}]{sonnendrucker1999semi}
Sonnendr{\"u}cker E, Roche J, Bertrand P and Ghizzo A (1999) {The
  semi-Lagrangian method for the numerical resolution of the Vlasov equation}.
\newblock \emph{Journal of Computational Physics} 149(2): 201--220.

\bibitem[{Verboncoeur(2005)}]{verboncoeur2005}
Verboncoeur JP (2005) {Particle simulation of plasmas: review and advances}.
\newblock \emph{Plasma Physics and Controlled Fusion} 47(5A): A231.

\bibitem[{Zweibel and Yamada(2009)}]{zweibel2009}
Zweibel E and Yamada M (2009) {Magnetic reconnection in astrophysical and
  laboratory plasmas}.
\newblock \emph{Annual Review of Astronomy and Astrophysics} 47: 291--332.

\end{thebibliography}
\bibliographystyle{sageh}

\appendix
\section{Appendix: Discontinuous Galerkin Poisson solver \label{app:dgsolver}}
First, we describe the 1d case which can then be easily extended to arbitrary dimensions, as we will show. The same polynomial space 
\begin{align*}
V^k_h = \{&\varphi \in L^2(\Omega): \varphi \in \mathbb{P}^k(I_i), \\ 
&I_i=[x_i,x_{i+1}], \quad i=0,\ldots,N-1\}, 
\end{align*}
that is used to perform the advections is chosen. As basis for $V_h^k$ the Lagrange polynomials interpolating on the $k+1$ Gauss--Legendre points $x_{il}$ restricted to the cell $I_i$ are chosen:
\[
 \varphi_{ij}(x) = \left\{ 
 \begin{aligned}
  \prod_{l=0,\ldots,k;l\neq j} \frac{x-x_{il}}{x_{ij}-x_{il}} & \quad x \in I_i \\
   0 \qquad \quad & \quad x \in \Omega \setminus I_i
 \end{aligned}
 \right. 
\]
The first step in deriving the discontinuous Galerkin Poisson solver consists in multiplying the Poisson equation \eqref{eq:poisson} by a test function $v\in V_h^k$, integrating over each cell $I_i = [x_i,x_{i+1}]$ and summing up all terms. For more details, see \cite{dgbook}. This gives, in 1d, the following equation
\begin{equation}
\begin{split}
\sum_{i=0}^{N-1} \bigg( \int_{I_i}&\phi'(x)v'(x)\,dx - [\phi'(x_{i})]\{v(x_i)\} - \\
&\{\phi'(x_{i})\}[v(x_i)] + \frac{\sigma}{h}[v(x_n)][\phi(x_n)] \bigg) \\ &= \int_{\Omega} (1-\rho(x))v(x)\,dx,
\end{split}
\label{eq:dgPoisson}
\end{equation}
where $\{v(x_i)\} = \tfrac{1}{2}(v(x_i^-) + v(x_i^+))$ is the average of the function $v$ at the point $x_i$ and $[v(x_i)] = v(x_i^-) - v(x_i^+) $ is the jump height. From the left hand side of \eqref{eq:dgPoisson} a symmetric bilinear form $a: V_h^k \times V_h^k \to \mathbb{R}$ can be derived. Therefore, we are looking for $\phi^{\text{DG}} \in V_h^k$ such that
\[
 a(\phi^{\text{DG}},v) = l(v) \quad \forall v \in V_h^k
\]
where the linear form $l$ is defined by the right-hand-side of \eqref{eq:dgPoisson}. Plugging $\phi^{\text{DG}}(x) = \sum_i\sum_{j=0}^k \phi(x_{ij}) \varphi_{ij}(x)$ and $v = \varphi_{lm}$ into $a$ results in a linear system of the form $Au=b$ where we are looking for the unknown $u$ specified by the coefficients $\phi(x_{ij})$. This method is called the symmetric interior penalty Galerkin (SIPG) method. It can be easily verified that this matrix $A$ is block tridiagonal, and that it can be constructed from two small matrices $M$ and $B$ of size $k+1$,
\[
A = 
\frac{1}{h}
\begin{pmatrix}
M   & B &   &         &     &     & B^T \\
B^T & M & B &         &     &     &     \\
    &   &   &         &     &     &     \\
    &   &   &         &     &     &     \\
    &   &   &         & B^T & M   & B   \\
B   &   &   &         &     & B^T & M   \\  
\end{pmatrix}
\]
To ensure coercivity of the bilinear form $a$ and positive definiteness of the matrix $A$, the condition $\sigma \geq k^2$ is required, see \cite{EPSHTEYN2007}. The matrices $M$ and $B$ are defined as follows

\begin{align*}
(M)_{i,j} =& +\tfrac{1}{2}\varphi_j'(0)\varphi_i(0) + \tfrac{1}{2}\varphi_j(0)\varphi_i'(0) +\sigma \varphi_j(0) \varphi_i(0) \\
&-\tfrac{1}{2}\varphi_j'(1)\varphi_i(1) - \tfrac{1}{2}\varphi_j(1)\varphi_i'(1) +\sigma \varphi_j(0) \varphi_i(1) \\
&+ \tfrac{1}{2} \sum_{l=0}^k \varphi_i'(x_l) \varphi_j'(x_l) w_l,  \\
&\\
(B)_{i,j} =& -\tfrac{1}{2}\varphi_j'(0)\varphi_i(1) + \tfrac{1}{2}\varphi_j(0)\varphi_i'(1) -\sigma \varphi_j(0) \varphi_i(1),
\end{align*}
where we denote with $\varphi_l(x)$ the Lagrange polynomials in the interval $[0,1]$ that interpolate the Gauss--Legendre quadrature points $x_l$. The $w_l$ are the Gaussian quadrature weights. In the higher dimensional setting, the bilinear form is given as follows
\begin{align*}
a(w,v) =& \sum_E \int_E \nabla v \cdot \nabla w - \sum_e \int_e \{\nabla v \cdot \textbf{n}_e\} [w] - \\
&\sum_e \int_e \{\nabla w \cdot n_e\}[v] + \sum_e \frac{\sigma}{|E|/|e|}\int_e [v][w],
\end{align*}
where $E$ in 2d are rectangles and in 3d are squares, and $e$ are the edges of the rectangles or the faces of the squares. Usually, the denominator in the last term is defined as $|e|^{\beta}$, but setting $\beta =1$ in 2d and $\beta =\tfrac{1}{2}$ in 3d is almost equivalent if the length of the edges do not vary much. Since in higher dimensions the basis is the tensor product of the 1d basis, the resulting matrix can be written in the following form in 2d
\[
 (h_y W_y \otimes A_x + h_x A_y \otimes W_x)u = b
\]
and in 3D
\[
\begin{aligned}
 (h_yh_z W_z \otimes W_y \otimes A_x + h_xh_z W_z \otimes A_y \otimes W_x + \\
 h_xh_y A_z \otimes W_y \otimes W_x)u = b,
\end{aligned}
\]
where $W_x$, $W_y$ and $W_z$ are diagonal matrices with the Gaussian quadrature weights on the diagonal.

To solve this linear system at every time step, we are using the conjugate gradient method. The reason why we prefer using this method over a direct method is, that as starting vector of this iterative method the electric potential $\phi$ computed at the previous time step can be used. Therefore, the number of iterations is small, since, if short time steps are used $\phi$ does not change dramatically. Moreover, the whole matrix is never constructed, a function is implemented which performs the action of this matrix applied on a vector. Thus, we use a completely matrix-free implementation.

If multiple compute nodes are used, only the boundary cells have to be transferred to the neighboring processes at each iteration. Therefore the method can be parallelized easily. Unfortunately, the matrix is singular since the system does not contain a constraint on the mean value of $\phi$. In principle this is not a problem since the electric field is the gradient of the potential function $\phi$ and therefore this constant shift vanishes. Nevertheless, it is difficult to construct a computationally cheap termination criteria in the CG method when the mean is growing continuously. Therefore, we embedded this condition in the CG method itself with almost no additional computational cost. The resulting CG method, which is mean preserving, is described in algorithm \ref{alg:CG}.
\begin{algorithm}[h]
\caption{Mean preserving conjugate gradient method}
\label{alg:CG}
\begin{algorithmic}
\STATE $r_0 = b - Au_0, \text{mean}(u_0)=0$
\STATE $r_0 = r_0 - \text{mean}(r_0)$, $p_0=r_0$
\STATE $\eta = (b+r_0)^Tu_0$
\WHILE {$i\leq \text{MAXITER}$} 
        \STATE $[w_i,\text{mean\textsubscript{w}}] = Ap_i$
        \STATE $\alpha_i = r_i^Tr_i/p_i^Tw_i$
        \STATE $u_{i+1} = u_{i} + \alpha_i p_i$
        \STATE $r_{i+1} = r_i - \alpha_i(w_i-\text{mean\textsubscript{w}})$
        \STATE $\eta = \eta + \alpha_i r_i^T r_i$
        \IF {$\|r_{i+1}\|_2 < \text{tol}\|b\|_2$}
            \STATE $\text{break}$
        \ENDIF
        \STATE $\beta_i = r_{i+1}^Tr_{i+1}/r^T_ir_i$
        \STATE $p_{i+1} = r_{i+1} + \beta_i p_i$
\ENDWHILE
\end{algorithmic}
\end{algorithm}

\begin{figure}[h]
\centering
\begin{subfigure}{.5\textwidth}
  \includegraphics[width=1.\linewidth]{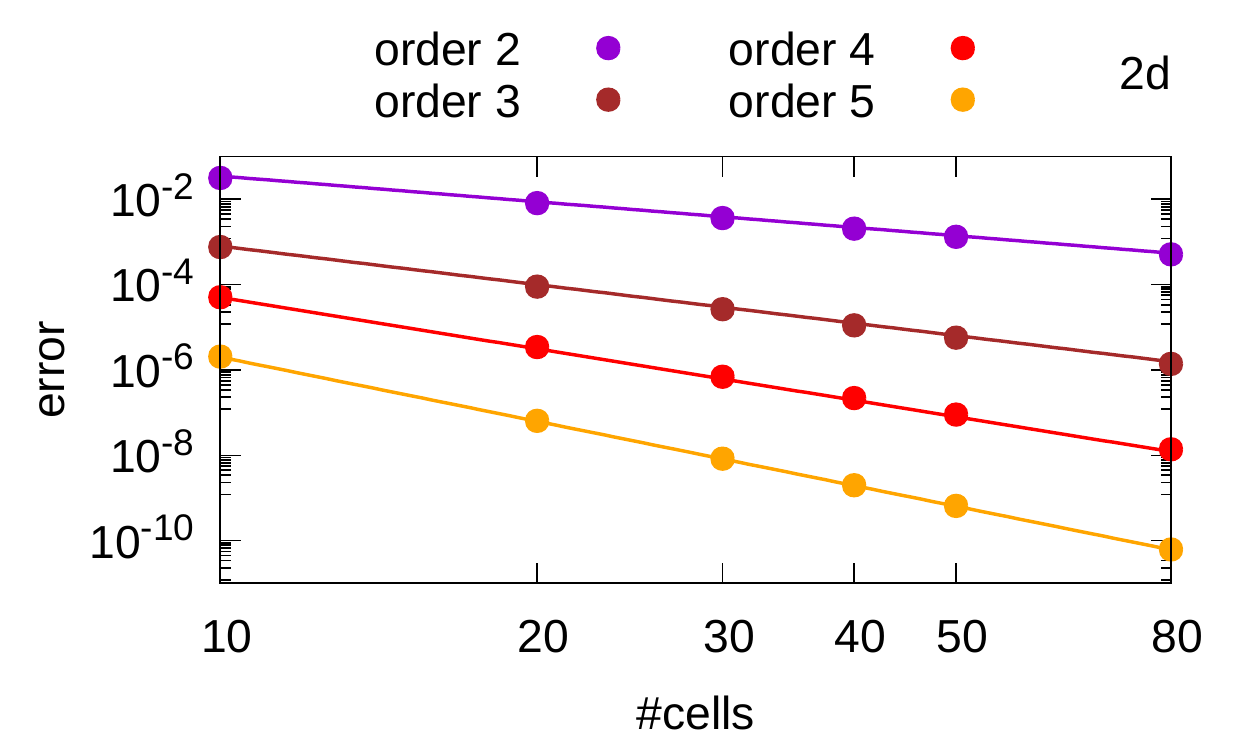}
\end{subfigure}%

\begin{subfigure}{.5\textwidth}
  \centering
  \includegraphics[width=1.\linewidth]{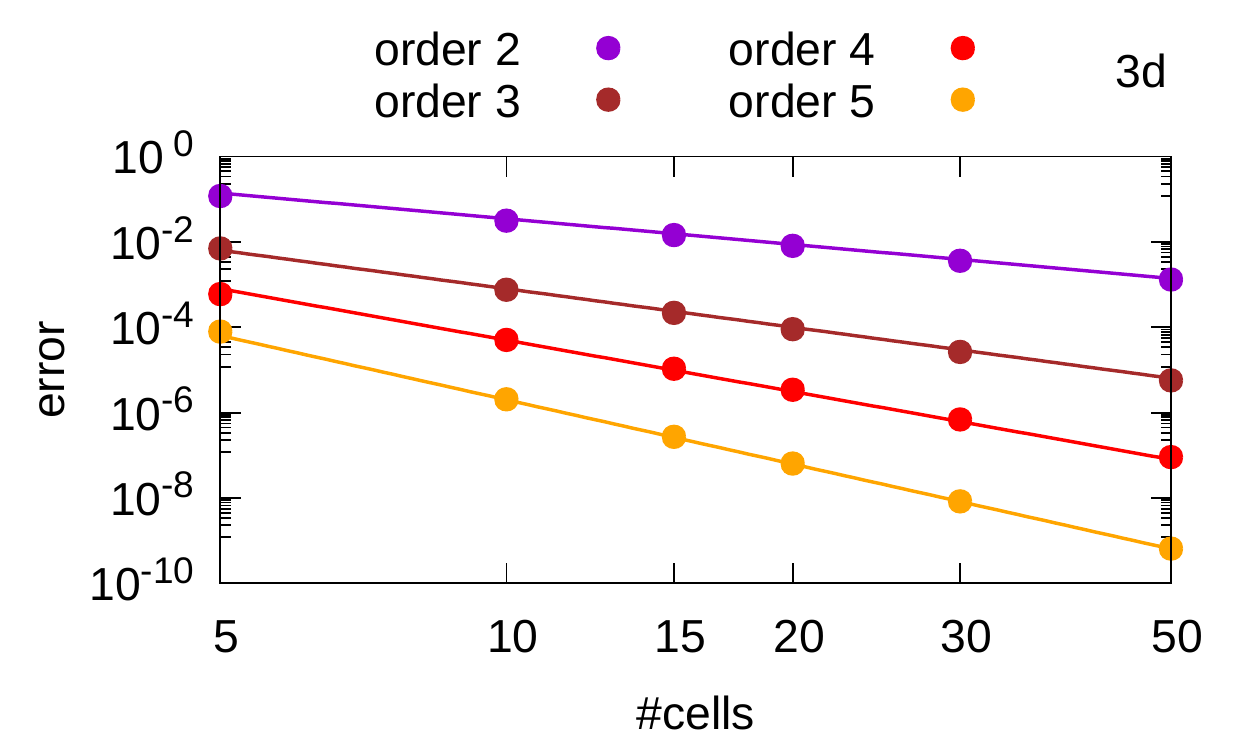}
\end{subfigure}
\caption{\label{fig:poisson2d3d}Relative $L^\infty$ error as a function of the number of cells for the discontinuous Galerkin Poisson solver in 2d and 3d.}
\end{figure}

To ensure that the dG Poisson solver is working properly, we test it first on a simple example with periodic boundary conditions,
\begin{align*}
 -\Delta \phi(x) &= \cos(0.5x).
\end{align*}
In figure~\ref{fig:poisson2d3d} we observe that the error is decreasing with order $k+1$ (note that the dG Poisson solver converges with the same order as the dG advection solver).

In the Vlasov-Poisson simulations, we set the relative tolerance in the CG method below $10^{-3}$ which is sufficient as can be seen in figures~\ref{fig:5dllbot} and~\ref{fig:6dllbot}. Moreover, $\sigma$ is set to its minimum allowed value. We observed, that when $\sigma$ is increasing, also the condition number of the matrix $A$ is increasing and more iterations are required to reach convergence. Additionally, we observe that convergence takes (almost) place within $(\prod_l^{dx} N_l(k_l+1))^{1/dx}$ iterations. Therefore, we set the maximum number of allowed iterations to this threshold times a safety factor.

\end{document}